\documentclass[nofootinbib]{revtex4}
\usepackage{graphicx}
\usepackage{dcolumn}
\usepackage{amsmath,amssymb,epsfig}
\usepackage{paralist}
\usepackage{comment}
\usepackage{wrapfig}
\usepackage{graphicx}
\usepackage{multirow}
\usepackage{color,soul}
\usepackage{suffix}
\usepackage{mathtools}
\usepackage[normalem]{ulem}

\newcommand\VTT[1]{\text{{\color[rgb]{0.75,0,0}~{#1}}}}
\WithSuffix\newcommand\VTT*[1]{{\bf{\color[rgb]{0.75,0,0}{#1}}}}

\definecolor{mysquiggly}{rgb}{0.9,0,0}

\makeatletter
\def\squiggly{\bgroup \markoverwith{\textcolor{mysquiggly}{\lower3.5\p@\hbox{{\sixly \char58}}}}\ULon}
\makeatother

\allowdisplaybreaks

\renewcommand{\vec}[1]{\boldsymbol{\mathrm{#1}}}

\begin{document}


\title{Wave-optical treatment of the shadow cast by a large gravitating sphere}

\author{Slava G. Turyshev$^{1}$, Viktor T. Toth$^2$
}

\affiliation{\vskip 3pt
$^1$Jet Propulsion Laboratory, California Institute of Technology,\\
4800 Oak Grove Drive, Pasadena, CA 91109-0899, USA
}%

\affiliation{\vskip 3pt
$^2$Ottawa, Ontario K1N 9H5, Canada
}%

\date{\today}

\begin{abstract}

We study the shadow cast by a large gravitating sphere, similar to our Sun. For this, we consider the gravitational field produced by a static mass monopole within the first post-Newtonian approximation of the general theory of relativity. We study the propagation of a monochromatic electromagnetic wave in the vicinity of a large, opaque, gravitating sphere. To treat the opaque nature of the body and its physical size, we implement fully absorbing boundary conditions and develop a wave-optical treatment of the shadow formed by this object. Based on this approach, we demonstrate that the structure of the shadow is determined by the Schwarzschild radius of the body and its physical size.  The shadow's boundary has the shape of a concave rotational hyperboloid bent inward due to the refractive properties of the curved spacetime. We show that there is no light in the shadow. However, even in the presence of gravity and related gravitational bending of photon trajectories there is the  bright spot of Arago that is formed behind a perfectly spherical obscuration.

\end{abstract}


\maketitle

\section{Introduction}
\label{sec:intro}

We investigate the scattering of light by the curvature of spacetime induced by the gravitational mass monopole of a large star. Specifically, we study the shadow cast by a large, massive, opaque sphere that is characterized by a weak gravitational field, similar to that of our Sun. This question is important for our ongoing efforts to study the solar gravitational lens (SGL) as the means for direct high-resolution imaging and spectroscopy of an exoplanet \cite{Turyshev:2017,Turyshev-Toth:2017,Turyshev-etal:2018,Turyshev-etal:2018-wp}.

Although  optical properties of the SGL are primarily determined by the solar Schwarzschild radius, the Sun itself, with its large physical dimensions, acts as a spherical obscuration that blocks a significant part of the incident wavefront.  It is known that in the absence of gravity,  a spherical obscuration with sharp boundaries placed in a beam of monochromatic light results in the formation of the Arago bright spot behind the sphere, which confirmed the wave-optical nature of light in a famous experiment.  How does gravity affect this process? Is there light in the shadow cast by a large gravitating sphere with sharp boundaries? These questions are of practical importance when considering the use of the SGL for imaging purposes, which was the primary motivation for this paper.

Towards this objective, recently we studied the electromagnetic (EM) field in the shadow cast by a large, opaque sphere with soft boundaries in flat spacetime \cite{Turyshev-Toth:2018}. We considered the scattering of a high-frequency  monochromatic EM wave by the large sphere. It is known that the presence of the spherical obscuration results in a geometric shadow of cylindrical shape behind the sphere. To describe this shadow, we developed a Mie theory that accounts for the obscuration and  demonstrated that there is indeed no EM field present in the shadow region. One exception is the presence of the bright spot of Arago, which can form at distances up to $z\ll kR_\odot^2$ (corresponding to a Fresnel number $F=kR_\odot^2/z \gg 1$), where $R_\odot$ is the solar radius and $k=2\pi/\lambda$ is the wavenumber corresponding to the observing wavelength. For optical and infrared wavelengths, i.e., for $\lambda\sim 1~\mu$m, these distances are up to $z\sim 98$~megaparsec (Mpc). In addition, to form the Arago spot the edge of the circular obscuration must be sufficiently smooth. However, it is not clear how the presence of gravity affects this process: if it only modifies the distance where such effect may exist or if it changes the entire phenomenon altogether.

It is known that gravity introduces refractive properties on spacetime \cite{Fock-book:1959}. As light propagates in the vicinity of the Sun, the direction of its propagation changes by the angle $\alpha=2r_g/b$, where $r_g=2GM/c^2$ is the Schwarzschild radius of the Sun, $M$ is the solar mass, and $b$ is the ray's solar impact parameter. Because of this refraction, in the solar vicinity photons propagate on hyperbolic trajectories that are bent towards the Sun, resulting in a specific shape of the shadow. Depending on the impact parameter, the rays will intersect at a heliocentric distance of $z\geq 547.8\,(b/R_\odot)^2$ astronomical units (AU) (see Fig.~\ref{fig:regions}).  As a result, the shadow forming behind a large, opaque, gravitating sphere, such is that of our Sun,  has a geometric boundary shaped like a concave rotational hyperboloid (studied in \cite{Turyshev-Toth:2017}).

In \cite{Turyshev-Toth:2017} we studied the properties of the caustic formed by the SGL at heliocentric distances beyond $z_0=547.8$~AU, where the interference region of the SGL begins (see Fig.~\ref{fig:regions}). This area has the shape of an elongated convex rotational hyperboloid whose outer boundary is set by the rays just grazing the Sun and thus intersecting the focal line at $z_0$ (i.e., defining the common vertex of the two hyperboloids that of the interference region and that of the shadow.)  This is the region  where the SGL acquires its most interesting properties, including extreme magnification and high optical resolution.  The region directly behind the Sun, corresponding to light rays with impact parameters $b\leq R^\star_\odot\equiv R_\odot+r_g$ is the region of the geometric shadow.  The third area is characterized by impact parameters $b>R^\star_\odot$ and two sets of heliocentric distances -- those smaller than $z_0$ and those larger than $z_0$, but outside the convex rotational hyperboloidal boundary of the interference region -- is called the region of geometric optics (see discussion in \cite{Deguchi-Watson:1986,Turyshev-Toth:2017}.)

The primary objective of this paper is to develop a wave-theoretical description of the shadow region of the Sun. This problem is characterized by a set of widely different scales, namely
\begin{inparaenum}[(i)]
\item the wavelength of radiation, $\lambda$, is much smaller than the radius of the Sun, $\lambda\ll R_\odot$, and
\item observational regions, at distances $r$ that much larger than the solar Schwarzschild radius, $r_g\ll R_\odot\ll r$.
\end{inparaenum}
This is exactly the situation encountered in the study of the optical properties of the SGL \cite{Turyshev:2017,Turyshev-Toth:2017} and its
applications. Such problems are typically tackled using a geometrical optics approximation or the usual methods of scalar optics \cite{Born-Wolf:1999}. These approximations can be avoided, however, within a wave-optical treatment, which is the topic of the present paper. In addition, such treatment allows us to study the wave nature of light in the presence of gravity, which may be important for some areas of astrophysics that deal with strong gravitational fields and/or large scales astronomical phenomena in the presence of gravity.

This paper is organized as follows: In Section~\ref{sec:em-waves-gr+pl}, we present the solution for the fictitious EM field of the solar shadow, given in terms of Debye potentials. Section~\ref{sec:shadow} is devoted to the investigation of the EM field in the solar shadow. Using the wave-optical treatment, we show that there is no light present in the shadow region of the SGL  to heliocentric distances up to 547.8~AU. In Section~\ref{sec:Arago}, we study the area on the optical axis directly behind the spherical obscuration in a search for the on-axis EM field that may appear there due to diffraction of light on the boundaries of a large sphere (that resembles the Sun with its physical size and mass, but has a sharp boundary.)   We show that even in the presence of gravity, there is the bright spot of Arago, that is formed in such conditions on axis behind a perfectly spherical obscuration. In Section~\ref{sec:disc}, we discuss the results that we obtained and present our conclusions.

\section{Electromagnetic waves in a static gravitational field}
\label{sec:em-waves-gr+pl}
\label{sec:maxwell}

To describe the optical properties of the SGL, we use a static, harmonic metric that represents the gravitational field produced by a mass monopole, taken in the post-Newtonian approximation of the general theory of relativity. Due to the spherical symmetry of the problem, we may introduce spherical coordinates $(r,\theta,\phi)$, where such a metric in the harmonic gauge is given by the following line element \cite{Fock-book:1959,Turyshev-Toth:2013}:
\begin{eqnarray}
ds^2&=&u^{-2}c^2dt^2-u^2\big(dr^2+r^2(d\theta^2+\sin^2\theta d\phi^2)\big),
\label{eq:metric-gen}
\end{eqnarray}
where, to the accuracy sufficient to describe light propagation in the solar system, the quantity $u$ has the form
\begin{eqnarray}
u=1+\frac{r_g}{2r}+{\cal O}(r_g^2),
\label{eq:w-PN}
\end{eqnarray}
with $r_g=2GM/c^2$ being the Schwarzschild radius of a gravitating body of mass $M$, thus representing the contribution of the mass monopole of the body's gravitational potential to the relativistic space-time (\ref{eq:metric-gen}).

\begin{figure}[t]
\includegraphics{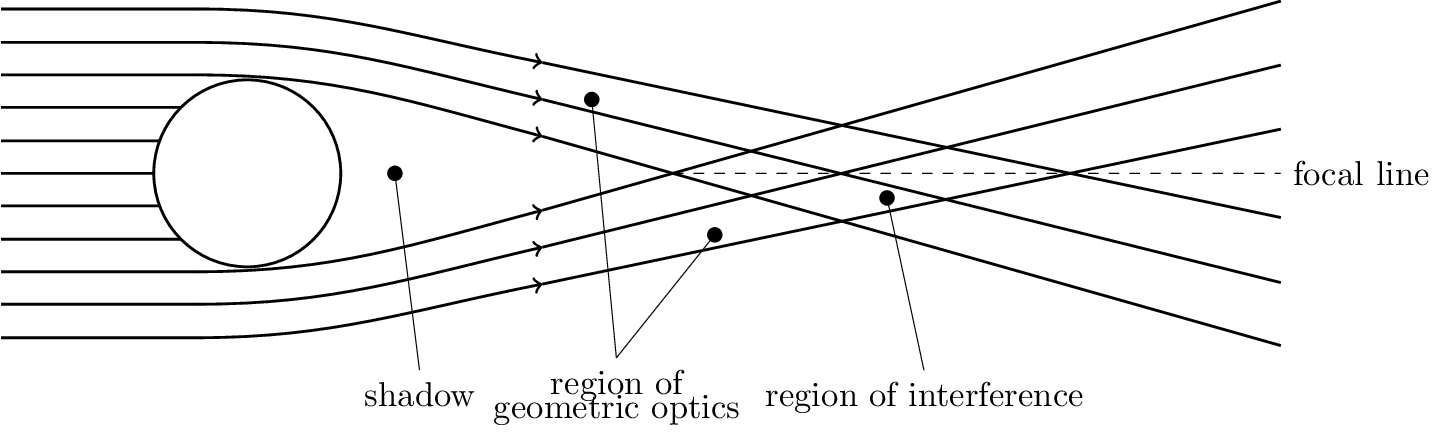}
\caption{
Three different regions of space associated with a monopole gravitational lens: the shadow, the region of geometric optics, and the region of interference (from \cite{Turyshev-Toth:2017}).\label{fig:regions}}
\end{figure}

Following \cite{Turyshev-Toth:2017}, we present Maxwell's equations in the spacetime given by (\ref{eq:metric-gen})--(\ref{eq:w-PN}) as
{}
\begin{eqnarray}
{\rm curl}\,{\vec D}&=&-  u^2\frac{1}{c}\frac{\partial \,{\vec B}}{\partial t}+{\cal O}(r_g^2),
\qquad ~{\rm div}\big(u^2\,{\vec D}\big)={\cal O}(r_g^2),
\label{eq:rotE_fl}\\[3pt]
{\rm curl}\,{\vec B}&=& u^2\frac{1}{c}\frac{\partial \,{\vec D}}{\partial t}+{\cal O}(r_g^2),
\qquad \quad
{\rm div }\big(u^2\,{\vec B}\big)={\cal O}(r_g^2),
\label{eq:rotH_fl}
\end{eqnarray}
where the differential operators ${\rm curl}$  and ${\rm div}$ are with respect to the 3-dimensional Euclidean flat metric.

As shown in \cite{Turyshev-Toth:2017}, following the Mie approach \cite{Mie:1908,Born-Wolf:1999}, one can separate the variables in the Maxwell equations (\ref{eq:rotE_fl})--(\ref{eq:rotH_fl}) and, by following the Mie approach, present a solution to these equations in the form of generic electric and magnetic Debye potentials, ${}^e{\hskip -1pt}\Pi$ and ${}^m{\hskip -1pt}\Pi$, respectively. Then, by matching these potentials to the incident Coulomb-modified plane wave, we find that both of these potentials in the vacuum may be given in terms of a single Debye potential $\Pi_0^{\rm g}(r, \theta)$:
\begin{align}
  \left( \begin{aligned}
{}^e{\hskip -1pt}\Pi& \\
{}^m{\hskip -1pt}\Pi& \\
  \end{aligned} \right) =&  \left( \begin{aligned}
\cos\phi \\
\sin\phi  \\
  \end{aligned} \right) \,\Pi_0^{\rm g}(r, \theta), & \hskip 2pt {\rm where}~~~~~
\Pi_0^{\rm g}(r, \theta)= E_0
\frac{u}{k^2r}\sum_{\ell=1}^\infty i^{\ell-1}\frac{2\ell+1}{\ell(\ell+1)}e^{i\sigma_\ell}
F_\ell(kr_g,kr) P^{(1)}_\ell(\cos\theta)+{\cal O}(r_g^2),
  \label{eq:Pi_ie*+}
\end{align}
where $k=2\pi/\lambda$ is the wavenumber of the EM wave, $F_\ell(kr_g,kr) $ and $\sigma_\ell$  are the Coulomb function and Coulomb phase shift of the $\ell$-th order, correspondingly \cite{Abramovitz-Stegun:1965,NIST-Handbook:2010,Morse-Feshbach:1953,Messiah:1968,Newton-book-2013}, with $P^{(1)}_\ell(\cos\theta)$ being the associated Legendre-polynomials of the first order. Note, that we use a heliocentric coordinate system with the $z$-axis oriented along the incident direction of the EM wave, where $\theta$ is the angle between the $z$-axis and the direction towards a particular observation point.

In the case of a spherically symmetric gravitational field, we may obtain the components of the EM field corresponding to a Debye potential $\Pi$ by first constructing the following quantities (see details in \cite{Turyshev-Toth:2017}):
\begin{eqnarray}
\alpha(r, \theta)&=&
-\frac{1}{u^2r^2}\frac{\partial}{\partial \theta}\Big[\frac{1}{\sin\theta} \frac{\partial}{\partial\theta}\big[\sin\theta\,(r\,\Pi)\big]\Big]=
\frac{1}{u}\Big\{\frac{\partial^2 }{\partial r^2}
\Big[\frac{r\,{\hskip -1pt}\Pi}{u}\Big]+k^2 \,u^4\Big[\frac{r\,{\hskip -1pt}\Pi}{u}\Big]\Big\},
\label{eq:alpha}\\
\beta(r, \theta)&=&\frac{1}{u^2r}
\frac{\partial^2 \big(r\,{\hskip -1pt}\Pi\big)}{\partial r\partial \theta}+\frac{ik\big(r\,{\hskip -1pt}\Pi\big)}{r\sin\theta},
\label{eq:beta}\\[0pt]
\gamma(r, \theta)&=&\frac{1}{u^2r\sin\theta}
\frac{\partial \big(r\,{\hskip -1pt}\Pi\big)}{\partial r}+\frac{ik}{r}
\frac{\partial\big(r\,{\hskip -1pt}\Pi\big)}{\partial \theta}.
\label{eq:gamma}
\end{eqnarray}
Using the  Debye potential $\Pi_0^{\rm g}(r, \theta)$  from (\ref{eq:Pi_ie*+}) in place of the quantity $\Pi$ in these definitions, we obtain the components of the EM field:
\begin{align}
  \left( \begin{aligned}
{  D}_r& \\
{  B}_r& \\
  \end{aligned} \right) =&  \left( \begin{aligned}
\cos\phi \\
\sin\phi  \\
  \end{aligned} \right) \,e^{-i\omega t}\alpha(r, \theta), &
    \left( \begin{aligned}
{  D}_\theta& \\
{  B}_\theta& \\
  \end{aligned} \right) =&  \left( \begin{aligned}
\cos\phi \\
\sin\phi  \\
  \end{aligned} \right) \,e^{-i\omega t}\beta(r, \theta), &
    \left( \begin{aligned}
{  D}_\phi& \\
{  B}_\phi& \\
  \end{aligned} \right) =&  \left( \begin{aligned}
-\sin\phi \\
\cos\phi  \\
  \end{aligned} \right) \,e^{-i\omega t}\gamma(r, \theta).
  \label{eq:DB-sol00p}
\end{align}

However, to derive the total EM field behind the Sun, we need to take into account the physical size of the opaque massive sphere, representing the Sun.
We can do that by imposing appropriate boundary conditions.  In this regard, the fully absorbing conditions that we used in \cite{Turyshev-Toth:2017,Turyshev-Toth:2018} are well-suited for this purpose. To impose such conditions, it is convenient  to represent the Coulomb function, $F_\ell$, in (\ref{eq:Pi_ie*+}) as a combination of two Hankel functions, $H^{(\pm)}_\ell$, given in the form $F_\ell=\big(H^{(+)}_\ell -H^{(-)}_\ell \big)/2i$. This yields the following expression for the Debye potential  $\Pi_0^{\rm g}(r,\theta)$ from (\ref{eq:Pi_ie*+}):
{}
\begin{eqnarray}
\Pi_0^{\rm g}(r, \theta)= -E_0
\frac{u}{2k^2r}\sum_{\ell=1}^\infty i^{\ell}\frac{2\ell+1}{\ell(\ell+1)}e^{i\sigma_\ell}
\Big(H^{(+)}_\ell(kr_g,kr) -H^{(-)}_\ell(kr_g,kr) \Big)P^{(1)}_\ell(\cos\theta)+{\cal O}(r_g^2).
  \label{eq:Pi_ie*+f}
\end{eqnarray}

Expression (\ref{eq:Pi_ie*+f}) allows us to set the boundary conditions on the surface of a fully absorbing sphere (see discussion in \cite{Turyshev-Toth:2017,Turyshev-Toth:2018}).  First, we recognize the asymptotic behavior of the Hankel functions $H^{(\pm)}_\ell(kr_g,kr)$. For large values of the argument $kr\rightarrow\infty $ (especially when $kr\gg \ell$, where $\ell$ is the order of the Coulomb function) and $r\gg r_{\rm t}=\sqrt{\ell(\ell+1)/k^2+r^2_g}-r_g$, these functions behave as \cite{Turyshev-Toth:2017}:
\begin{eqnarray}
\lim_{kr\rightarrow\infty} H^{(\pm)}_\ell(kr_g,kr)
&\sim&\exp\Big[\pm i\Big(kr+kr_g\ln2kr-\frac{\pi\ell}{2}+\sigma_\ell+\frac{\ell(\ell+1)}{2kr}\Big)\Big]+{\cal O}\big((kr)^{-3}\big).
\label{eq:FassC}
\end{eqnarray}
Thus, the radial function $H^{(+)}_\ell(kr_g,kr)$ represents the outgoing wave, while $H^{(-)}_\ell(kr_g,kr)$ is the incoming wave. This observation emphasizes the fact, demonstrated by the structure of (\ref{eq:Pi_ie*+f}), that in the absence of any interaction, the Debye potential of a free EM wave may be thought of as a superposition of incoming and outgoing waves.

Next, we recognize that the smallest possible impact parameter represents light rays that are grazing the solar surface, which corresponds to the heliocentric distance of $R_\odot^*=R_\odot+r_g$. This extra $r_g$ term, in addition to the solar radius, accounts for the fact that rays of light are bent by the gravitational field of the mass monopole even as they approach the source. In other words, the rays with $b\leq R^*_\odot$ will hit the Sun and will be absorbed by its surface. Therefore, we require that for impact parameters $b\leq R_\odot^*$ there will be no scattered or coherently retransmitted waves. To implement this condition, we rely on a relation between index $\ell$ that, in a semiclassical sense, is analogous to the partial momentum of a particle, and the impact parameter $b$, given as $\ell\simeq kb$ \cite{Glauber-Matthiae:1970,Sharma-Somerford:1990,Sharma-Sommerford-book:2006,Friedrich-book-2006}. With this semiclassical relation, our boundary condition is equivalent to the requirement that the Sun completely absorbs waves with partial momenta $\ell \leq\ell_{\rm max}=kR_\odot^*$. This fully absorbing boundary condition is easy to implement using the solution (\ref{eq:Pi_ie*+f}) and subtracting the part that corresponds to the outgoing wave, which is blocked by the sphere.

As a result, the solution for the entire Debye potential, $\Pi (r, \theta)$, in the area behind the sphere (i.e, for $|\theta|\leq {\textstyle\frac{\pi}{2}}$), in the presence of gravity, takes the following form \cite{Turyshev-Toth:2017}:
{}
\begin{eqnarray}
\Pi (r, \theta)&=&\Pi_0^{\rm g} (r, \theta)+\delta \Pi^{\rm g} (r, \theta)=\Pi_0^{\rm g} (r, \theta)+E_0\frac{u}{2k^2r}\sum_{\ell=1}^{kR_\odot^*} i^{\ell}\frac{2\ell+1}{\ell(\ell+1)}e^{i\sigma_\ell} H^{(+)}_\ell(kr_g,kr) P^{(1)}_\ell(\cos\theta),
\label{eq:P_g-tot+q}
\end{eqnarray}
where $\Pi_0^{\rm g} (r, \theta)$ is the Debye potential of the incident wave given by (\ref{eq:Pi_ie*+})  and $\delta \Pi^{\rm g} (r, \theta)$ is the shadow potential.

In  Ref.~\cite{Turyshev-Toth:2017}, we obtained a closed form analytical solution for the Debye potential, $\Pi_0^{\rm g}(r, \theta)$, from (\ref{eq:Pi_ie*+}), in the form
{}
\begin{equation}
\Pi^{\rm g}_0(\vec r)=-\psi_0\frac{iu}{k}\frac{1-\cos\theta}{\sin\theta}\Big(e^{ikz}{}_1F_1[1+ikr_g,2,ikr(1-\cos\theta)]-e^{ikr}{}_1F_1[1+ikr_g,2,2ikr]\Big)+{\cal O}(r_g^2),
\label{eq:sol-Pi0}
\end{equation}
giving the Debye potential of the incident wave in terms of the confluent hypergeometric function, ${}_1F_1[a,b,z]$ \cite{Herlt-Stephani:1975,Herlt-Stephani:1976,Turyshev-Toth:2017}. This solution is always finite and is valid for any angle $\theta$. The availability of this exact, closed form solution for the Debye potential allowed us to investigate the structure of the caustic formed by the lens. In addition, the analytical solution given in (\ref{eq:sol-Pi0}) was used \cite{Turyshev-Toth:2017} to derive the EM field in the interference region of the SGL, including the EM field on the image plane and the corresponding Poynting-vector. As both forms are equivalent, for various calculations where  $\Pi_0^{\rm g}(r, \theta)$ plays an important role, we will use either the form (\ref{eq:Pi_ie*+}) for this potential or that given by (\ref{eq:sol-Pi0}).

Also, by taking into account the asymptotic behavior of the hypergeometric functions ${}_1F_1[1+ikr_g,2,ikr(1-\cos\theta)]$ and  ${}_1F_1[1+ikr_g,2,2ikr]$ at large values of the argument $k(r-z)\gg1$ \cite{Abramovitz-Stegun:1965} (and, thus, for $\theta\gg 1/\sqrt{kr}>0$), we may establish the asymptotic behavior of the closed form analytical solution for the Debye potential $\Pi^{\rm g}_0(r,\theta)$ from (\ref{eq:sol-Pi0}). The corresponding expression takes the following form \cite{Turyshev-Toth:2017} :
{}
\begin{eqnarray}
\Pi^{\rm g}_0(r,\theta)&=&E_0\frac{u}{k^2r\sin\theta}\Big\{e^{ik\big(z-r_g\ln k(r-z)\big)}
- e^{ik\big(r+r_g\ln k(r-z)\big)+2i\sigma_0}-\nonumber\\
&-&{\textstyle\frac{1}{2}}(1-\cos\theta)\Big(
e^{-ik\big(r+r_g\ln 2kr\big)}
- e^{ik\big(r+r_g\ln 2kr\big)+2i\sigma_0}\Big)
+{\cal O}\Big(\frac{ikr_g^2}{r-z}\Big)\Big\},
\label{eq:Pi-ass*}
\end{eqnarray}
were $z=r\cos\theta$, constant $\sigma_0$ is given as $\sigma_0=\arg \Gamma(1-ikr_g)$ \cite{Abramovitz-Stegun:1965,NIST-Handbook:2010}, which for large values of $kr_g\rightarrow\infty$ behaves  as  \cite{Turyshev-Toth:2017}
\begin{eqnarray}
e^{2i\sigma_0}=
\frac{\Gamma(1-ikr_g)}{\Gamma(1+ikr_g)}=e^{-2ikr_g\ln (kr_g/e)-i\frac{\pi}{2}}\Big(1+{\cal O}((kr_g)^{-1})\Big).
\label{eq:Gam_rat}
\end{eqnarray}

The second term in (\ref{eq:P_g-tot+q}), $\delta \Pi^{\rm g} (r, \theta)$, is the Debye potential responsible for the ``solar shadow'',  given as
{}
\begin{eqnarray}
\delta \Pi^{\rm g} (r, \theta)=
E_0\frac{u}{2k^2r}\sum_{\ell=1}^{kR_\odot^*} i^{\ell}\frac{2\ell+1}{\ell(\ell+1)}e^{i\sigma_\ell} H^{(+)}_\ell(kr_g,kr) P^{(1)}_\ell(\cos\theta),
\label{eq:P_g-shdw}
\end{eqnarray}
which was obtained as a result of applying the fully absorbing boundary conditions, as discussed above. The fully absorbing boundary is, thus, the asymptotic boundary condition set on the future light cone and deals with the fact that the physical size of the opaque Sun is much larger than its Schwarzschild radius, $R_\odot\gg r_g$.  Clearly, $\delta \Pi^{\rm g} (r, \theta)$  is also a solution of the Maxwell equations  (\ref{eq:rotE_fl})--(\ref{eq:rotH_fl}) with the corresponding EM field computed with  (\ref{eq:alpha})--(\ref{eq:gamma}), (\ref{eq:DB-sol00p}).

Eqs.~(\ref{eq:P_g-tot+q})--(\ref{eq:P_g-shdw}) capture all aspects of the EM field behind a large gravitating opaque sphere. This solution is valid at all distances and angles. However, because of its complexity, numerical methods are required to explore its physical implications. In this paper, we explore the solution in the far field \cite{Kerker-book:1969}, where several approximation methods are feasible.
Specifically, to study the contribution of the shadow potential (\ref{eq:P_g-shdw}), it is convenient to use the asymptotic behavior of the Hankel functions (\ref{eq:FassC}) and present this potential in its asymptotic form as
\begin{eqnarray}
\delta \Pi^{\rm g} (r, \theta)&=&E_0\frac{u}{2k^2r}e^{ik(r+r_g\ln 2kr)}\sum_{\ell=1}^{kR_\odot^*} \frac{2\ell+1}{\ell(\ell+1)}e^{i\big(2\sigma_\ell+\frac{\ell(\ell+1)}{2kr}\big)} P^{(1)}_\ell(\cos\theta)+{\cal O}(r_g^2).
\label{eq:de_P-g}
\end{eqnarray}
Note that, as $r_g\ll R_\odot $, the asymptotic from of the solution (\ref{eq:de_P-g}) is valid right outside the Sun, for distances $r\geq R^*_\odot$ and improves as distance increases. As a result, the solution for the total Debye potential (\ref{eq:P_g-tot+q}), takes the  form:
{}
\begin{eqnarray}
\Pi (r, \theta)&=&
\Pi_0^{\rm g} (r, \theta)+E_0\frac{u}{2k^2r}e^{ik(r+r_g\ln 2kr)}\sum_{\ell=1}^{kR_\odot^*} \frac{2\ell+1}{\ell(\ell+1)}e^{i\big(2\sigma_\ell+\frac{\ell(\ell+1)}{2kr}\big)}
P^{(1)}_\ell(\cos\theta).
\label{eq:P_g-tot}
\end{eqnarray}

The behavior of $\Pi_0^{\rm g} (r, \theta)$ given by (\ref{eq:sol-Pi0}) was established in \cite{Turyshev-Toth:2017} and now is well understood. It is finite and computable for any relevant distances and angles, including forward scattering at $\theta=0$. We used this solution in \cite{Turyshev-Toth:2017} to investigate the EM field in the image plane situated in the interference region of the SGL beyond 547.8 AU (Fig.~\ref{fig:regions}). In that region and near the optical axis $\theta\approx 0$, the EM field is completely given by (\ref{eq:Pi_ie*+}) (or, equivalently, by (\ref{eq:sol-Pi0})), which we used to derive the field in the image plane and study the optical properties of the SGL. Expression $\delta \Pi^{\rm g} (r, \theta)$ from (\ref{eq:de_P-g}) is important, as it provides access to the EM field in the shadow region, which is the main subject of this paper.

We have thus obtained the Debye potential (\ref{eq:P_g-tot}) representing the total solution for the problem of the scattering of EM waves by the gravitational field of a large spherical star, similar to our Sun. Specifically (\ref{eq:P_g-tot}), together with (\ref{eq:alpha})--(\ref{eq:DB-sol00p}), represents the solution for the EM field propagating on the background of a spherically symmetric, static gravitational field produced by a large star with opaque surface. This solution allows us to study the physical behavior of the EM field in the three regions behind this star (see Fig.~\ref{fig:regions}), namely
\begin{inparaenum}[(i)]
\item the shadow region, where no incident light enters (i.e., for impact parameters $0\leq b\leq R_\odot^*$),
\item the region of geometrical optics, where only one ray passes through any given point (i.e., for impact parameters $b>R_\odot^*$ and heliocentric distances $r< R_\odot^2/2r_g$), and
\item the interference region (i.e., $r\geq R_\odot^2/2r_g$, especially for forward scattering, $\theta\approx 0$).
\end{inparaenum}
The interference region formed by solar gravity beyond 547.8~AU behind the Sun is of the greatest importance. The EM field in region is fully characterized by the Debye potential (\ref{eq:sol-Pi0}), which was investigated in  \cite{Turyshev-Toth:2017} including the description of the EM field the region of geometrical optics. Here we focus our attention on the shadow region of the Sun, where the shadow potential (\ref{eq:P_g-shdw}) also plays critical role.

\section{EM field in the shadow region}
\label{sec:shadow}

To study the EM field in the shadow region, we need to express the  components of the EM field in terms of the variables involved. We do that by using the expression (\ref{eq:P_g-shdw})  for the Debye potential $\delta \Pi^{\rm g} (r, \theta)$ responsible for the solar shadow and deriving the components of the  fictitious EM field produced by this potential. This field is generated by $\delta \Pi^{\rm g} (r, \theta)$ to compensate the incident EM wave for a particular impact parameter  $0\simeq b \leq R^\star_\odot$ of the incident EM wave.
To develop analytical expressions for this fictitious field, we use (\ref{eq:de_P-g}) in the expressions (\ref{eq:alpha})--(\ref{eq:gamma}), deriving the factors $\alpha(r,\theta), \beta(r,\theta)$ and $\gamma(\theta)$, which, to ${\cal O}\big(r^2_g\big)$, have the form
{}
\begin{eqnarray}
\alpha(r,\theta) &=& E_0\frac{e^{ik(r+r_g\ln 2kr)}}{k^2r^2}\sum_{\ell=1}^{kR_\odot^*}(\ell+{\textstyle\frac{1}{2}})e^{i\big(2\sigma_\ell+\frac{\ell(\ell+1)}{2kr}\big)}P^{(1)}_\ell(\cos\theta) \Big\{u^2-\frac{ikr_g}{\ell(\ell+1)} \Big\},
  \label{eq:alpha*1*}\\
\beta(r,\theta) &=& -E_0\frac{ue^{ik(r+r_g\ln 2kr)}}{ikr}\sum_{\ell=1}^{kR^*_\odot}\frac{\ell+{\textstyle\frac{1}{2}}}{\ell(\ell+1)}e^{i\big(2\sigma_\ell+\frac{\ell(\ell+1)}{2kr}\big)} \Big\{\frac{\partial P^{(1)}_\ell(\cos\theta)}
{\partial \theta}\Big(1-\frac{\ell(\ell+1)}{2k^2r^2u^2}+\frac{ir_g}{2kr^2}\Big)+\frac{P^{(1)}_\ell(\cos\theta)}{\sin\theta} \Big\},
  \label{eq:beta*1*}\\
\gamma(r,\theta) &=& -E_0\frac{ue^{ik(r+r_g\ln 2kr)}}{ikr}\sum_{\ell=1}^{kR_\odot^*}\frac{\ell+{\textstyle\frac{1}{2}}}{\ell(\ell+1)}e^{i\big(2\sigma_\ell+\frac{\ell(\ell+1)}{2kr}\big)}\Big\{\frac{\partial P^{(1)}_\ell(\cos\theta)}
{\partial \theta}+\frac{P^{(1)}_\ell(\cos\theta)}{\sin\theta}\Big(1-
\frac{\ell(\ell+1)}{2k^2r^2u^2}+\frac{ir_g}{2kr^2}\Big)
 \Big\}.~~~~~
  \label{eq:gamma*1*}
\end{eqnarray}

Note that, to derive these expressions, we relied on the asymptotic behaviour of the Hankel functions (\ref{eq:FassC}), which were developed up to ${\cal O}((kr)^{-3})$. Therefore, (\ref{eq:alpha*1*})--(\ref{eq:gamma*1*}) are also valid to the same order in their phase term, $\exp\big(i(2\sigma_\ell+{\ell(\ell+1)}/{2kr})\big)$. If needed, using the approach presented in Appendix B of \cite{Turyshev-Toth:2018-plasma}, we may include terms with higher powers in $(kr)^{-1}$ in (\ref{eq:FassC}) and, thus, extend the order of approximation  in (\ref{eq:alpha*1*})--(\ref{eq:gamma*1*}).

Together with (\ref{eq:DB-sol00p}), these expressions describe the fictitious EM field produced by the solar shadow. They contain all the relevant information about the opaque nature of the solar surface and the  size of the Sun.
We rely on (\ref{eq:alpha*1*})--(\ref{eq:gamma*1*}) to investigate the EM field in the region of interest: the solar shadow.

\subsection{Solution for the function $\alpha(r,\theta)$ and the radial components of the EM field}
\label{sec:radial-comp}

We begin with the investigation of $\alpha(r,\theta)$, given by (\ref{eq:alpha*1*}). To evaluate this expression in the shadow region for large angles, $\theta \gg \sqrt{2r_g/r}$, we use the appropriate asymptotic representation for $P^{(1)}_l(\cos\theta)$  \cite{Bateman-Erdelyi:1953,Korn-Korn:1968,Kerker-book:1969}:
{}
\begin{align}
P^{(1)}_\ell(\cos\theta)  &=
\dfrac{-\ell}{\sqrt{2\pi \ell \sin\theta}}\Big(e^{i(\ell+\frac{1}{2})\theta+i\frac{\pi}{4}}+e^{-i(\ell+\frac{1}{2})\theta-i\frac{\pi}{4}}\Big)+{\cal O}(\ell^{-\textstyle\frac{3}{2}}) ~~~~~\textrm{for}~~~~~ 0<\theta<\pi.
\label{eq:P1l<}
\end{align}
which results in the following form of  (\ref{eq:alpha*1*}):
{}
\begin{eqnarray}
\alpha(r,\theta) &=& -E_0\frac{e^{ik(r+r_g\ln 2kr)}}{k^2r^2}\sum_{\ell=1}^{kR_\odot^*}\frac{(\ell+{\textstyle\frac{1}{2}})\sqrt{\ell}}{\sqrt{2\pi\sin\theta}}\Big(u^2-\frac{ikr_g}{\ell(\ell+1)} \Big)e^{i\big(2\sigma_\ell+\frac{\ell(\ell+1)}{2kr}\big)}\Big(e^{i(\ell+\frac{1}{2})\theta+i\frac{\pi}{4}}+e^{-i(\ell+\frac{1}{2})\theta-i\frac{\pi}{4}}\Big).
  \label{eq:P_sum}
  \end{eqnarray}

We recognize that for small impact parameters, $0\leq b\leq r_g$, light rays will be captured by the solar Schwarzschild field and, thus, will not be transmitted.  For large impact parameters, $\ell\geq kr_g$, we may replace $\ell+1\rightarrow \ell$ and $\ell+\textstyle\frac{1}{2}\rightarrow \ell$ and also replace the sum in (\ref{eq:P_sum}) with an integral:
{}
\begin{eqnarray}
\alpha(r,\theta) &=& -E_0\frac{e^{ik(r+r_g\ln 2kr)}}{k^2r^2}\int_{\ell=1}^{kR_\odot^*}\frac{\ell\sqrt{\ell}d\ell}{\sqrt{2\pi\sin\theta}}\Big(u^2-\frac{ikr_g}{\ell^2} \Big)e^{i\big(2\sigma_\ell+\frac{\ell^2}{2kr}\big)}\Big(e^{i(\ell\theta+\frac{\pi}{4})}+e^{-i(\ell\theta+\frac{\pi}{4})}\Big),
  \label{eq:Pi_s_exp1}
  \end{eqnarray}
and evaluate this integral by the method of stationary phase. We recall that this method allows evaluating integrals of the following type:
{}
\begin{equation}
I=\int A(\ell)e^{i\varphi(\ell)}d\ell, \qquad
\ell\in\mathbb{R},
\label{eq:stp-1}
\end{equation}
where the amplitude $A(\ell)$ is a slowly varying function of $\ell$, while $\varphi(\ell)$ is a rapidly varying function of $\ell$.
The integral (\ref{eq:stp-1}) may be replaced, to good approximation, with a sum over the points of stationary phase, $\ell_0\in\{\ell_{1,2,..}\}$, for which $d\varphi/d\ell=0$. Defining $\varphi''=d^2\varphi/d\ell^2$, we obtain the integral
{}
\begin{equation}
I\simeq\sum_{\ell_0\in\{\ell_{1,2,..}\}} A(\ell_0)\sqrt{\frac{2\pi}{\varphi''(\ell_0)}}e^{i\big(\varphi(\ell_0)+{\textstyle\frac{\pi}{4}}\big)}.
\label{eq:stp-2}
\end{equation}

Expression (\ref{eq:Pi_s_exp1}) shows that the $\ell$-dependent part of the phase has the following structure:
{}
\begin{equation}
\varphi_{\pm}(\ell)=\pm\big(\ell\theta+{\tfrac{\pi}{4}}\big)+\frac{\ell^2}{2kr}+2\sigma_\ell+{\cal O}\big((kr)^{-3}\big).
\label{eq:S-l}
\end{equation}

We recall that the Coulomb phase shift $\sigma_\ell$  has the form \cite{Abramovitz-Stegun:1965,Turyshev-Toth:2017}:
{}
\begin{equation}
\sigma_\ell=\sigma_0-\sum_{j=1}^\ell\arctan\frac{kr_g}{j}, \qquad \sigma_0=\arg\Gamma(1-ikr_g),
\label{eq:S-l-g-s*}
\end{equation}
where $\sigma_0$ was evaluated in \cite{Turyshev-Toth:2017} to be
{}
\begin{eqnarray}
\sigma_0&=& -kr_g\ln \frac{kr_g}{e}-\frac{\pi}{4}.
\label{eq:sig-0*}
\end{eqnarray}

We may replace the sum in (\ref{eq:S-l-g-s*}) with an integral and, for $r_g\ll b\leq R^\star_\odot$ and thus, $kr_g\ll \ell\leq kR^\star_\odot$, evaluate $\sigma_\ell$ as:
{}
\begin{eqnarray}
\sigma_\ell&=& -kr_g\ln \ell +{\cal O}(r_g^2).
\label{eq:sig-l*}
\end{eqnarray}

Therefore, the points of stationary phase, where $d\varphi_{\pm}/d\ell=0$, are given from (\ref{eq:S-l}) by the following equation:
{}
\begin{equation}
\pm\theta= 2\arctan \frac{kr_g}{\ell}-\frac{\ell}{kr},
\label{eq:S-l-pri-g*}
\end{equation}
which is the equation for families of hyperbolae. If we take, from the semiclassical approximation, the connection between the partial momentum, $\ell$ and the impact parameter, $b$, given as $\ell\simeq kb,$ then for small angles $\theta$ (or, large distances from the sphere, $r\gg R_\odot$),  from (\ref{eq:S-l-pri-g*}) we see that the points of stationary phase  must satisfy the equation  (see \cite{Turyshev-Toth:2017} for details):
{}
\begin{equation}
\frac{1}{r}=\pm\frac{\sin\theta}{b}+\frac{2r_g}{b^2} + {\cal O}(r_g^2),
\label{eq:theta-b0*}
\end{equation}
within the shadow region characterized by $r_g\leq b\leq R_\odot^\star$.

\begin{figure}[t]
\vspace{-10pt}
\includegraphics[width=0.48\linewidth]{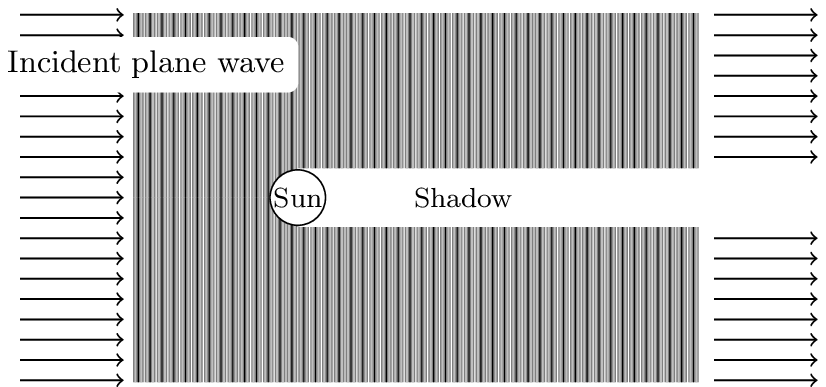}
\hskip 10pt
\includegraphics[width=0.48\linewidth]{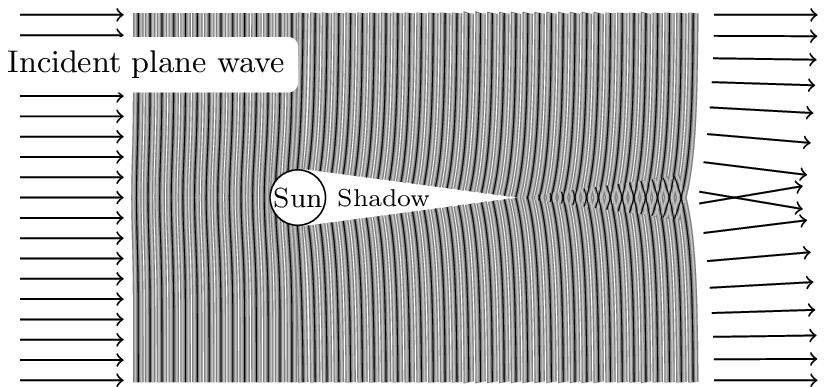}
\caption{Diffraction of light on a large spherical opaque obscuration. Left: Geometric shadow region with a cylindrical shape, formed in flat space-time (see details in \cite{Turyshev-Toth:2018}).  Right: shadow with a hyperbolic shape (no rays of light exist in this region), caustic formed at the interference region, and the region of geometric optics in the case of a large gravitating sphere. The arrows indicate the direction of the Poynting-vector, i.e., the direction of wavefront propagation.
\label{fig:shadow}}
\vspace{0pt}
\end{figure}

The presence of the last term in (\ref{eq:theta-b0*}) defines the properties of the shadow region behind the Sun. This term is absent in flat spacetime, where $r_g=0$ and the boundary is set by the straight lines $b=r\sin\theta$.  Based on the sum in (\ref{eq:Pi_s_exp1}), the largest impact parameter that defines the shadow region  is $b^{\rm max}=R_\odot+r_g$, thus, in any given plane that contains the focal line, (\ref{eq:theta-b0*}) is bounded by the two most extreme hyperbolae describing the boundary that coincides with the two outmost rays that are almost grazing, but are still absorbed by the Sun on opposite sides in the forward direction, $0\leq\theta\leq \frac{\pi}{2}$ and, thus, setting the hyperboloid shape of the boundary of the shadow.

The result given by (\ref{eq:theta-b0*}) is different from that given in \cite{Turyshev-Toth:2018} because, in the case of light bent by gravity, individual photon trajectories are bent towards the gravitating body, causing light rays to enter the cylindrical region of geometric shadow that is prohibited in the case of a flat space-time studied in \cite{Turyshev-Toth:2018} (see Fig.~{\ref{fig:shadow}).

Equation (\ref{eq:theta-b0*}) yields two families of solutions for  the points of stationary phase:
{}
\begin{equation}
\ell^{(1)}_{0}= \mp kr\Big(\sin\theta+\frac{2r_g}{r}\frac{1}{\sin\theta}\Big)+{\cal O}(r_g^2), \qquad {\rm and} \qquad
\ell^{(2)}_{0}= \pm \frac{2kr_g}{\sin\theta}+{\cal O}(r_g^2).
\label{eq:S-l-pri}
\end{equation}

The `$\pm$' or `$\mp$' signs in (\ref{eq:S-l-pri}) represent the families of fictitious ``shadow'' rays propagating on opposite sides from the Sun. Also, the two families of solutions represent two different waves. Thus, the family $\ell^{(1)}_{0}$ represents the incident wave. The family $\ell^{(2)}_{0}$ describes the scattered wave (see Fig.~1 in \cite{Turyshev-Toth:2017} and relevant discussion therein and also in \cite{Deguchi-Watson:1986,Turyshev-Toth:2017}).

The first family of solutions of (\ref{eq:S-l-pri}), given by $\ell^{(1)}_0$,  allows us to compute the phase for the points of stationary phase (\ref{eq:S-l}) for the EM waves moving towards the interference region (a similar calculation was done in \cite{Turyshev-Toth:2018-plasma}):
{}
\begin{equation}
\varphi_{\pm}(\ell^{(1)}_0)=\pm\textstyle{\frac{\pi}{4}}-\textstyle{\frac{1}{2}}\theta^2 kr-kr_g\ln kr(1-\cos\theta)-kr_g\ln2kr+{\cal O}(kr\theta^4,kr_g\theta^2).
\label{eq:S-l2g}
\end{equation}

Computing the second derivative of the phase, $\varphi (\ell)$, given by (\ref{eq:S-l}), with respect to $\ell$, we have
{}
\begin{equation}
\varphi''_{\pm}(\ell)=\frac{1}{kr}+\frac{2kr_g}{\ell^2}.
\label{eq:S-l22}
\end{equation}
After substituting here $\ell^{(1)}_0$ from (\ref{eq:S-l-pri}), we have
\begin{eqnarray}
\varphi''(\ell^{(1)}_0)\equiv \dfrac{d^2\varphi_{\pm}}{d\ell^2} \Big|_{\ell=\ell^{(1)}_0} &=&
\dfrac{1}{kr}\Big(1+\frac{2r_g}{r\sin^2\theta}+{\cal O}(r^2_g)\Big)
\qquad  \Rightarrow \qquad
\sqrt{\frac{2\pi}{\varphi''(\ell_0)}}=\sqrt{2\pi kr}\Big(1-\frac{r_g}{r\sin^2\theta}+{\cal O}(r^2_g)\Big).~~~
\label{eq:S-l2}
\end{eqnarray}

Using (\ref{eq:S-l2}), we may compute the amplitude of the integrand in (\ref{eq:Pi_s_exp1}), for $\ell_0=\ell^{(1)}_0$, which takes the form
{}
\begin{eqnarray}
A(\ell_0)\sqrt{\frac{2\pi}{\varphi''(\ell_0)}}&=&\frac{\ell_0\sqrt{\ell_0}}{\sqrt{2\pi \sin\theta}}
\Big(u^2-\frac{ikr_g}{\ell^2_0} \Big)\sqrt{\frac{2\pi}{\varphi''(\ell_0)}}=
(\mp1)^{\textstyle\frac{3}{2}} k^2r^2\sin\theta\Big(1+\frac{2r_g}{r\sin^2\theta}\big(1-\frac{i}{2kr}\big)+{\cal O}(r_g^2,\frac{r_g}{r}\theta^2)\Big).~~~
\label{eq:S-l3p+*}
\end{eqnarray}
Because of its smallness, we can drop the $\propto 1/(kr)$ term in the parentheses of this expression.

As a result, for the first family of solutions, $\ell_0^{(1)}$, from (\ref{eq:S-l-pri}),  the expression for $\alpha(r,\theta)$  from (\ref{eq:Pi_s_exp1}) corresponding to the fictitious incident wave takes the form
{}
\begin{eqnarray}
\alpha^{\rm in}_\pm(r,\theta)&=&-E_0u^{-1}\sin\theta \Big(1+\frac{r_g}{r(1-\cos\theta)}+{\cal O}(\theta^4,\frac{r_g}{r}\theta^2)\Big)e^{i\big(kr\cos\theta-kr_g\ln kr(1-\cos\theta)\big)}.
\label{eq:Pi_s_exp4+1*}
\end{eqnarray}

Now we consider the second family of solutions in (\ref{eq:S-l-pri}), given by $\ell^{(2)}_{0}$, which leads to the following expression for the stationary phase
{}
\begin{eqnarray}
\varphi_{\pm}(\ell^{(2)}_0)&=&
\pm\textstyle{\frac{\pi}{4}}-kr_g\ln 2kr+kr_g\ln (1-\cos\theta)-2kr_g\ln \frac{kr_g}{e}+{\cal O}(kr_g \theta^2).
\label{eq:S-l27*+}
\end{eqnarray}
Using this result, from (\ref{eq:Pi_s_exp1}) we compute the phase of the corresponding solution (by combining the relativistic phase and the $\ell$-dependent contribution):
{}
\begin{eqnarray}
\varphi^{(2)}_{\pm}(r,\theta)&=&kr+kr_g\ln 2kr+\varphi^{(2)}_{\pm}(\ell_0)+\textstyle{\frac{\pi}{4}}=
k(r+r_g\ln kr(1-\cos\theta))+2\sigma_0+{\cal O}(kr_g \theta^2).
\label{eq:S-l27}
\end{eqnarray}

Now, using (\ref{eq:S-l22}) and $\ell^{(2)}_{0}$ from (\ref{eq:S-l-pri}), we compute the second derivative of the phase with respect to $\ell$:
{}
\begin{equation}
\varphi''_{\pm}(\ell_0)=\frac{1}{kr}+\frac{\sin^2\theta}{2kr_g}\simeq\frac{\sin^2\theta}{2kr_g}\Big(1+\frac{2r_g}{r\sin^2\theta}\Big)+{\cal O}(\theta^5), \qquad {\rm thus, } \qquad
\sqrt{\frac{2\pi}{\varphi''(\ell_0)}}=\frac{\sqrt{4\pi kr_g}}{\sin\theta}\Big(1-\frac{r_g}{r\sin^2\theta}\Big)+{\cal O}(\theta^5).
\label{eq:S-l2202}
\end{equation}

At this point, we may evaluate the amplitude of the integrand in (\ref{eq:Pi_s_exp1}), for $\ell_0=\ell^{(2)}_0$, which takes the form
{}
\begin{eqnarray}
A(\ell_0)\sqrt{\frac{2\pi}{\varphi''(\ell_0)}}&=&\frac{\ell_0\sqrt{\ell_0}}{\sqrt{2\pi \sin\theta}} \Big(u^2-\frac{ikr_g}{\ell^2_0} \Big)\sqrt{\frac{2\pi}{\varphi''(\ell_0)}}=
(\mp1)^{\textstyle\frac{3}{2}} \frac{4k^2r^2_g}{\sin^3\theta}\Big(1-\frac{r_g}{r\sin^2\theta}-\frac{i\sin^2\theta}{4kr_g}\Big).~~~~
\label{eq:S-l3p+*=}
\end{eqnarray}
As a result,  the expression for $\delta\alpha(r,\theta)$, for the scattered wave from (\ref{eq:Pi_s_exp1}) for $\ell^{(2)}_0$ and for $\theta\gg\sqrt{2r_g/r}$, takes the form
{}
\begin{eqnarray}
\alpha^{\rm sc}_\pm(r,\theta)&=&-E_0\Big(\frac{2r_g}{r}\Big)^2\frac{1}{\sin^3\theta} e^{ik(r+r_g\ln kr(1-\cos\theta)+2i\sigma_0}
\sim {\cal O}(r_g^2).
\label{eq:alpha-2}
\end{eqnarray}
Thus, to the accepted approximation, there is no scattered wave in the radial direction. This result is consistent with that reported in \cite{Turyshev-Toth:2017}.

The results (\ref{eq:Pi_s_exp4+1*}) and (\ref{eq:alpha-2}) are the radial components of the EM wave  (\ref{eq:DB-sol00p}), $D_r, B_r$, corresponding to the two families of impact parameters given  by (\ref{eq:S-l-pri}). We use these solutions to determine the resulting EM field in the shadow region.

\subsection{Evaluating the function $\beta(r,\theta)$}
\label{sec:amp_func-beta}

To investigate the behavior $\beta(r,\theta)$ from (\ref{eq:beta*1*}), we neglect terms of ${\cal O}(r_g/kr^2)$ and obtain the following  expression for $\beta(r,\theta)$:
{}
\begin{eqnarray}
\beta(r,\theta) &=& -E_0\frac{ue^{ik(r+r_g\ln 2kr)}}{ikr}\sum_{\ell=1}^{kR^*_\odot}\frac{\ell+{\textstyle\frac{1}{2}}}{\ell(\ell+1)}e^{i\big(2\sigma_\ell+\frac{\ell(\ell+1)}{2kr}\big)} \Big\{\frac{\partial P^{(1)}_\ell(\cos\theta)}
{\partial \theta}\Big(1-\frac{\ell(\ell+1)}{2k^2r^2u^2}\Big)+\frac{P^{(1)}_\ell(\cos\theta)}{\sin\theta}
 \Big\}.
  \label{eq:beta**1+}
\end{eqnarray}

To evaluate the magnitude of the function $\beta(r, \theta)$, we  need to establish the asymptotic behavior of $P^{(1)}_{l}(\cos\theta)/\sin\theta$ and $\partial P^{(1)}_{l}(\cos\theta)/\partial \theta$. For fixed $\theta$ and $kr_g\leq\ell\rightarrow\infty$, this behavior is given as (this can be obtained directly from (\ref{eq:P1l<})):
{}
\begin{eqnarray}
\frac{P^{(1)}_\ell(\cos\theta)}{\sin\theta}
&=& \Big(\frac{2\ell}{\pi\sin^3\theta}\Big)^{\frac{1}{2}} \sin\Big((\ell+{\textstyle\frac{1}{2}})\theta-{\textstyle\frac{\pi}{4}}\Big)+{\cal O}(\ell^{-\textstyle\frac{3}{2}}),
\label{eq:pi-l*}\\
\frac{dP^{(1)}_\ell(\cos\theta)}{d\theta}
&=&  \Big(\frac{2\ell^3}{\pi\sin\theta}\Big)^{\frac{1}{2}} \cos\Big((\ell+{\textstyle\frac{1}{2}})\theta-{\textstyle\frac{\pi}{4}}\Big)+{\cal O}(\ell^{-\textstyle\frac{1}{2}}).
\label{eq:tau-l*}
\end{eqnarray}

With these approximations, the function $\beta(r,\theta)$ in the shadow region but outside the optical axis takes the form:
{}
\begin{eqnarray}
\beta(r,\theta) &=& -E_0\frac{ue^{ik(r+r_g\ln 2kr)}}{ikr}\sum_{\ell=1}^{kR^*_\odot}\frac{\ell+{\textstyle\frac{1}{2}}}{\ell(\ell+1)}e^{i\big(2\sigma_\ell+\frac{\ell(\ell+1)}{2kr}\big)}\times\nonumber\\
&&\hskip 10pt \times\,
\Big\{\Big(\frac{2\ell^3}{\pi\sin\theta}\Big)^{\frac{1}{2}}\Big(1-\frac{\ell(\ell+1)}{2k^2r^2u^2}\Big)\cos\Big((\ell+{\textstyle\frac{1}{2}})\theta-{\textstyle\frac{\pi}{4}}\Big)+ \Big(\frac{2\ell}{\pi\sin^3\theta}\Big)^{\frac{1}{2}} \sin\Big((\ell+{\textstyle\frac{1}{2}})\theta-{\textstyle\frac{\pi}{4}}\Big)
 \Big\}.
  \label{eq:S1-v0s}
\end{eqnarray}

For large $\ell\gg kr_g$, the first term in the curly brackets in (\ref{eq:S1-v0s}) dominates, so this expression may be given as
{}
\begin{eqnarray}
\beta(r,\theta) &=& -E_0\frac{ue^{ik(r+r_g\ln 2kr)}}{ikr}\sum_{\ell=1}^{kR^*_\odot}\frac{\ell+{\textstyle\frac{1}{2}}}{\ell(\ell+1)}
\Big(\frac{2\ell^3}{\pi\sin\theta}\Big)^{\frac{1}{2}}\Big(1-\frac{\ell(\ell+1)}{2k^2r^2u^2}\Big)
e^{i\big(2\sigma_\ell+\frac{\ell(\ell+1)}{2kr}\big)}
\cos\Big((\ell+{\textstyle\frac{1}{2}})\theta-{\textstyle\frac{\pi}{4}}\Big).
  \label{eq:S1-v0s+}
\end{eqnarray}

To evaluate $\beta(r,\theta)$ from (\ref{eq:S1-v0s+}), we again use the method of stationary phase. For this, representing (\ref{eq:S1-v0s+}) in the form of an integral over $\ell$, we have
{}
\begin{eqnarray}
\beta(r,\theta) &=& E_0\frac{ue^{ik(r+r_g\ln 2kr)}}{kr}\int_{\ell=1}^{kR^*_\odot}\frac{\sqrt{\ell}d\ell}{\sqrt{2\pi\sin\theta}}
\Big(1-\frac{\ell^2}{2k^2r^2u^2}\Big)e^{i\big(2\sigma_\ell+\frac{\ell(\ell+1)}{2kr}\big)}
\Big(e^{i(\ell\theta+{\textstyle\frac{\pi}{4}})}-e^{-i(\ell\theta+{\textstyle\frac{\pi}{4}})}\Big).
\label{eq:S1-v0s+int*}
\end{eqnarray}

Equation (\ref{eq:S1-v0s+int*}) shows that the $\ell$-dependent part of the phase have a structure identical to (\ref{eq:S-l}). Therefore, the same solutions for the points of stationary phase apply. Then, with (\ref{eq:S-l2}), for the first family of solutions $\ell^{(1)}_0$  (\ref{eq:S-l-pri}), from (\ref{eq:S1-v0s+int*}) we have
{}
\begin{eqnarray}
A(\ell_0)\sqrt{\frac{2\pi}{\varphi''(\ell_0)}}&=& \frac{\sqrt{\ell}}{\sqrt{2\pi\sin\theta}}\Big(1-\frac{\ell^2}{2k^2r^2u^2}\Big)\sqrt{\frac{2\pi}{\varphi''(\ell_0)}}=\sqrt{\mp1} kru^{-2}
\Big\{1-{\textstyle\frac{1}{2}}\theta^2-\frac{r_g}{r}+{\cal O}(\theta^4, \frac{r_g}{r}\theta^2)\Big\}=\nonumber\\
&=&
\pm \sqrt{\mp1}kru^{-1}\Big\{\cos\theta-\frac{r_g}{r}+
{\cal O}(\theta^4, \frac{r_g}{r}\theta^2)\Big\}.~~~
\label{eq:S-l3p}
\end{eqnarray}

As a result, similarly to (\ref{eq:Pi_s_exp4+1*}), the expression for the function $ \beta_\pm(r,\theta)$  of the fictitious incident wave takes the form
{}
\begin{eqnarray}
 \beta^{\rm in}_\pm(r,\theta)&=& \pm\sqrt{\mp1}E_0u^{-1}\Big\{\cos\theta-\frac{r_g}{r}+
{\cal O}(\theta^4, \frac{r_g}{r}\theta^2)\Big\}e^{i\big(\pm{\textstyle{\frac{\pi}{4}}}+kr\cos\theta-kr_g\ln kr(1-\cos\theta)\big)+i{\textstyle{\frac{\pi}{4}}}}
=\nonumber\\
&=&-E_0u^{-1}\Big(\cos\theta-\frac{r_g}{r}\Big) e^{ik\big(r\cos\theta-r_g\ln kr(1-\cos\theta)\big)}
+{\cal O}(\theta^4, \frac{r_g}{r}\theta^2).~~~~~
\label{eq:Pi_s_exp4+1pp}
\end{eqnarray}

Now we turn our attention to the second family of solutions in (\ref{eq:S-l-pri}).
Similarly to (\ref{eq:S-l3p+*=}), for $\ell^{(2)}_0$, we have
{}
\begin{eqnarray}
A(\ell_0)\sqrt{\frac{2\pi}{\varphi''(\ell_0)}}&=&\frac{\sqrt{\ell}}{\sqrt{2\pi\sin\theta}} \Big(1-\frac{\ell^2}{2k^2r^2}\Big)\sqrt{\frac{2\pi}{\varphi''(\ell_0)}}=
\sqrt{\pm1} \frac{kr_g}{2\sin^2\textstyle{\frac{1}{2}}\theta}+{\cal O}(\theta^4,r^2_g),~~~~
\label{eq:S-l3p+*=2}
\end{eqnarray}
which yields the following result for $ \beta^{[0]}_\pm(r,\theta)$:
{}
\begin{eqnarray}
 \beta^{\rm sc}_\pm(r,\theta)&=& -\sqrt{\pm1} E_0\frac{r_g}{2r\sin^2\textstyle{\frac{1}{2}}\theta}e^{i\big(\pm\textstyle{\frac{\pi}{4}}-\textstyle{\frac{\pi}{4}}+kr+kr_g\ln kr(1-\cos\theta) +2\sigma_0\big)}+{\cal O}(\theta^4,r^2_g)
=\nonumber\\
&=&-E_0\frac{r_g}{2r\sin^2\textstyle{\frac{1}{2}}\theta}e^{i\big(k(r+r_g\ln kr(1-\cos\theta))+2\sigma_0\big)}
+{\cal O}(\theta^4, \frac{r_g}{r}\theta^2).~~~~~
\label{eq:beta-2}
\end{eqnarray}

With  (\ref{eq:DB-sol00p}), expressions (\ref{eq:Pi_s_exp4+1pp}) and (\ref{eq:beta-2}) provide the form of the function $\beta(r,\theta)$ that determines the $D_\theta, B_\theta$ components  of the fictitious EM field representing the solar shadow.

\subsection{Evaluating the  function $\gamma(r,\theta)$}
\label{sec:amp_func-der}

To determine the remaining components of the EM field (\ref{eq:DB-sol00p}), we need to evaluate  the behavior of the function $\gamma(r,\theta)$ from (\ref{eq:gamma*1*}) that is given in the following from:
{}
\begin{eqnarray}
\gamma(r,\theta) &=& -E_0\frac{ue^{ik(r+r_g\ln 2kr)}}{ikr}\sum_{\ell=1}^{kR_\odot^*}\frac{\ell+{\textstyle\frac{1}{2}}}{\ell(\ell+1)}e^{i\big(2\sigma_\ell+\frac{\ell(\ell+1)}{2kr}\big)} \Big\{\frac{\partial P^{(1)}_\ell(\cos\theta)}
{\partial \theta}+\frac{P^{(1)}_\ell(\cos\theta)}{\sin\theta}\Big(1-
\frac{\ell(\ell+1)}{2k^2r^2u^2}\Big) \Big\},
  \label{eq:gamma**1}
\end{eqnarray}
where, similarly to (\ref{eq:beta**1+}), we dropped the insignificant $r_g/kr^2$ term.

To evaluate this expression, we use the asymptotic behavior of $P^{(1)}_{l}(\cos\theta)/\sin\theta$ and $\partial P^{(1)}_{l}(\cos\theta)/\partial \theta$ given by (\ref{eq:pi-l*}) and (\ref{eq:tau-l*}), correspondingly, and rely on the method of stationary phase. Similarly to (\ref{eq:S1-v0s}), we drop the second term in the curly brackets in (\ref{eq:gamma**1}). The remaining expression for $\gamma(r, \theta)$ is now determined by the following integral:
{}
\begin{eqnarray}
\gamma(r,\theta) &=& E_0\frac{ue^{ik(r+r_g\ln 2kr)}}{kr}\int_{\ell=1}^{kR_\odot^*}\frac{\sqrt{\ell}}{\sqrt{2\pi\sin\theta}}e^{i\big(2\sigma_\ell+\frac{\ell(\ell+1)}{2kr}\big)}
\Big(e^{i(\ell\theta+{\textstyle\frac{\pi}{4}})}-e^{-i(\ell\theta+{\textstyle\frac{\pi}{4}})}\Big).
  \label{eq:gamma**1*}
\end{eqnarray}

Clearly, this expression yields the same points of the stationary phase (\ref{eq:S-l}) and, thus, all the relevant results obtained in Sec.~\ref{sec:radial-comp} apply. Therefore, the $\ell$-dependent amplitude of (\ref{eq:gamma**1*}), is evaluated for $\ell^{(1)}_0$ from (\ref{eq:S-l-pri}) as
{}
\begin{equation}
A(\ell_0)\sqrt{\frac{2\pi}{\varphi''(\ell_0)}}=
\frac{\sqrt{\ell}}{\sqrt{2\pi\sin\theta}}\sqrt{\frac{2\pi}{\varphi''(\ell_0)}}=\pm\sqrt{\mp1}
kr+{\cal O}(\theta^3, \frac{r_g}{r}\theta^2).
\label{eq:S-l3p+0*}
\end{equation}

The function $ \gamma_\pm(r,\theta)$ of the incident wave is then given as
{}
\begin{eqnarray}
 \gamma^{\rm in}_\pm(r,\theta)&=&\pm \sqrt{\mp1}E_0u
e^{i\big(\pm{\textstyle{\frac{\pi}{4}}}+kr\cos\theta-kr_g\ln kr(1-\cos\theta)\big)\big)+i{\textstyle{\frac{\pi}{4}}}}=
-E_0u
e^{i\big(kr\cos\theta-kr_g\ln kr(1-\cos\theta)\big)}+{\cal O}(\theta^4,\frac{r_g}{r}\theta^2).~~
\label{eq:Pi_s_exp4+1gg}
\end{eqnarray}

For the second family of solutions (\ref{eq:S-l-pri}), we get the following result:
{}
\begin{equation}
A(\ell_0)\sqrt{\frac{2\pi}{\varphi''(\ell_0)}}=
\frac{\sqrt{\ell}}{\sqrt{2\pi\sin\theta}}\sqrt{\frac{2\pi}{\varphi''(\ell_0)}}=\pm\sqrt{\pm1} \frac{kr_g}{2\sin^2\textstyle{\frac{1}{2}}\theta}+{\cal O}(\theta^4,r^2_g),
\label{eq:S-l3p+0*2}
\end{equation}
which yields a result for $ \gamma^{\rm sc}_\pm(r,\theta)$ that is identical to (\ref{eq:beta-2}):
{}
\begin{eqnarray}
 \gamma^{\rm sc}_\pm(r,\theta)&=&
-E_0\frac{r_g}{2r\sin^2\textstyle{\frac{1}{2}}\theta}e^{i\big(k(r+r_g\ln kr(1-\cos\theta))+2\sigma_0\big)}
+{\cal O}(\theta^4, \frac{r_g}{r}\theta^2).~~~~~
\label{eq:gamma-2}
\end{eqnarray}

Again, with  (\ref{eq:DB-sol00p}), expressions (\ref{eq:Pi_s_exp4+1gg}) and (\ref{eq:gamma-2}) provide the function $\gamma(r,\theta)$ that determines the $D_\phi, B_\phi$ components  of the fictitious EM field representing the solar shadow.

At this point, we have all the necessary ingredients to present the solution for the  EM field in the shadow region.

\subsection{Solution for the EM field in the shadow outside the focal axis}
\label{sec:EM-shadow}

To determine the components of the EM field, we use the expressions that we obtained for the functions $\alpha(r,\theta)$, $\beta(r,\theta)$ and $\gamma(r,\theta)$, which are given by (\ref{eq:Pi_s_exp4+1*}), (\ref{eq:Pi_s_exp4+1pp}) and (\ref{eq:Pi_s_exp4+1gg}), correspondingly, and substitute them in (\ref{eq:DB-sol00p}). As a result, we establish the fictitious solution for the incident EM field in the shadow region for angles $\theta\geq\sqrt{2r_g/r}$:
{}
\begin{eqnarray}
\left( \begin{aligned}
{    D}_r^{\rm in}& \\
{    B}_r^{\rm in}& \\
  \end{aligned} \right) &=&  -E_0u^{-1}\left( \begin{aligned}
\cos\phi \\
\sin\phi  \\
  \end{aligned} \right)
\Big(1+\frac{r_g}{r(1-\cos\theta)}\Big)e^{i\big(kr\cos\theta-kr_g\ln kr(1-\cos\theta)-\omega t\big)}
+{\cal O}(\theta^4, \frac{r_g}{r}\theta^2),
  \label{eq:DB-r*+*p2}\\
  \left( \begin{aligned}
{   D}^{\rm in}_\theta& \\
{   B}^{\rm in}_\theta& \\
  \end{aligned} \right) &=&
  -E_0u^{-1}
\left( \begin{aligned}
\cos\phi \\
\sin\phi  \\
  \end{aligned} \right)\Big(\cos\theta-\frac{r_g}{r}\Big) e^{ik\big(r\cos\theta-r_g\ln kr(1-\cos\theta)-\omega t\big)}
+{\cal O}(\theta^4, \frac{r_g}{r}\theta^2),
  \label{eq:DB-t-pl=p1}\\
\left( \begin{aligned}
{   D}^{\rm in}_\phi& \\
{   B}^{\rm in}_\phi& \\
  \end{aligned} \right) &=&
 - E_0u
\left( \begin{aligned}
-\sin\phi \\
\cos\phi  \\
  \end{aligned} \right) e^{i\big(kr\cos\theta-kr_g\ln kr(1-\cos\theta)-\omega t\big)}+{\cal O}(\theta^4,\frac{r_g}{r}\theta^2).
  \label{eq:DB-t-pl=p2}
\end{eqnarray}

At the same time, the EM field produced by the Debye potential $\Pi_0$ of the incident wave (\ref{eq:Pi-ass*}) has the form \cite{Turyshev-Toth:2017}:
{}
\begin{eqnarray}
  \left( \begin{aligned}
{    D}^{\rm (0)}_r& \\
{    B}^{\rm (0)}_r& \\
  \end{aligned} \right) &=& E_0u^{-1}
\left( \begin{aligned}
\cos\phi \\
\sin\phi  \\
  \end{aligned} \right)\sin\theta \Big(1+\frac{r_g}{r(1-\cos\theta)}\Big)\,e^{i\big(k(r\cos\theta-r_g\ln kr(1-\cos\theta))-\omega t\big)},
  \label{eq:DB-t-pl=p10} \\
    \left( \begin{aligned}
{    D}^{\rm (0)}_\theta& \\
{    B}^{\rm (0)}_\theta& \\
  \end{aligned} \right) &=& E_0u^{-1}
\left( \begin{aligned}
\cos\phi \\
\sin\phi  \\
  \end{aligned} \right) \Big(\cos\theta-\frac{r_g}{r}\Big)\,e^{i\big(k(r\cos\theta-r_g\ln kr(1-\cos\theta))-\omega t\big)},
   \label{eq:DB-th=p10} \\
   \left( \begin{aligned}
{    D}^{\rm (0)}_\phi& \\
{    B}^{\rm (0)}_\phi& \\
  \end{aligned} \right) &=&E_0u
  \left( \begin{aligned}
-\sin\phi \\
\cos\phi  \\
  \end{aligned} \right) \,e^{i\big(k(r\cos\theta-r_g\ln kr(1-\cos\theta))-\omega t\big)}.
  \label{eq:DB-t-pl=p20}
\end{eqnarray}

The total field, in accord with (\ref{eq:P_g-tot+q}), is given by the sum of (\ref{eq:DB-t-pl=p1})--(\ref{eq:DB-t-pl=p2}) and (\ref{eq:DB-t-pl=p10})--(\ref{eq:DB-t-pl=p20}). It is easy to see that this total EM field in the shadow behind the Sun completely vanishes.
However, tis is not yet a complete story: we recall that in the case when gravity is involved, there are two waves that characterize the scattering process, namely the incident wave (\ref{eq:DB-t-pl=p10})--(\ref{eq:DB-t-pl=p20}) and the scattered wave, which was computed in \cite{Turyshev-Toth:2017} (see equations (49)--(50) therein) and is given as
{}
\begin{eqnarray}
    \left( \begin{aligned}
{    D}^{\rm (0)}_\theta& \\
{    B}^{\rm (0)}_\theta& \\
  \end{aligned} \right)_{\rm \hskip -4pt s} =   \left( \begin{aligned}
{    B}^{\rm (0)}_\phi& \\
-{    D}^{\rm (0)}_\phi& \\
  \end{aligned} \right)_{\rm \hskip -4pt s}= E_0
\left( \begin{aligned}
\cos\phi \\
\sin\phi  \\
  \end{aligned} \right) \frac{r_g}{2r\sin^2\frac{\theta}{2}}\,e^{i\big(k(r+r_g\ln kr(1-\cos\theta))+2\sigma_0-\omega t\big)},
  \qquad
    \left( \begin{aligned}
{    D}^{\rm (0)}_r& \\
{    B}^{\rm (0)}_r& \\
  \end{aligned} \right)_{\rm \hskip -4pt s} &=& {\cal O}(r^2_g).~~~
   \label{eq:scat-th}
\end{eqnarray}

To verify that the scattered field also vanishes in the shadow, we need to compute the total scattered EM field in the shadow region behind the Sun. We do this similarly to the discussion of the incident field above, by adding the corresponding components of the scattered field (\ref{eq:scat-th}) and those of fictitious scattered fields induced by the solar shadow (similar to the approach discussed in  \cite{Borovoi:2013}). Using the results for the functions $\alpha^{\rm sc}_\pm(r,\theta)$, $\beta^{\rm sc}_\pm(r,\theta)$ and $\gamma^{\rm sc}_\pm(r,\theta)$, given  by (\ref{eq:alpha-2}), (\ref{eq:beta-2}) and (\ref{eq:gamma-2}), correspondingly, we compute the components of the fictitious scattered fields ${\vec D}_{\rm sc}$ and ${\vec B}_{\rm  sc}$ by substituting these functions in (\ref{eq:DB-sol00p}).  Then, with the help of (\ref{eq:P_g-tot+q}) we compute the total EM field behind the Sun as
${\vec D}_{\rm s}={\vec D}^{(0)}_{\rm s}+{\vec D}_{\rm sc}$ and ${\vec B}_{\rm s}={\vec B}^{(0)}_{\rm s}+{\vec B}_{\rm sc}$. This exercise allows us to verify that there is no scattered wave behind the Sun, as expected.

This completes our investigation of the EM field in the solar geometric shadow region. We found that, in the presence of gravity, the solar shadow occupies the heliocentric distances $R_\odot\leq r\leq R^2/2r_g\simeq 547.8$~AU and, because of the gravitational deflection of light,  it has the shape of a rotational hyperboloid whose boundary is determined by (\ref{eq:theta-b0*}). We found no coherently re-transmitted or scattered light in the solar geometric shadow region, which is expected from  the fully absorbing boundary conditions that we implemented in Sec.~\ref{sec:maxwell} precisely for this purpose.

\section{EM field on the optical axis and the bright spot of Arago}
\label{sec:Arago}

For practical applications of the SGL, we are interested in the EM field in the area directly behind the Sun \cite{Turyshev-etal:2018}. It is known that, in flat spacetime, the bright spot of Arago is formed exactly on the optical axis  behind an obscuration with a perfectly spherical boundary. This effect is a well-known manifestation of the wave nature of light. It is due the diffraction of light waves on the edges or boundaries of the obscuration. We search for such an effect in the presence of gravity, taking into account the related gravitational bending of light trajectories.

In flat spacetime, the bright spot of Arago forms under conditions consistent with the Fresnel approximation \cite{Born-Wolf:1999} which are satisfied within the range of distances $z \ll k R^2_\odot$ from the Sun,  corresponding to heliocentric distances of $z \ll 98\,(1~\mu{\rm m}/\lambda)$~Mpc. One encounters this spot behind the sphere on the axis having the amplitude of the incident wave. Once we take gravity into account, our region of interest is confined to the shadow region, extending from the solar surface to 547.8~AU. 

It is also known that the Fresnel diffraction mechanism responsible for the Arago spot is affected by the roughness of the spherical boundary. For the Sun to exhibit such an effect, its surface roughness (the width of the Fresnel zone surrounding the solar disk \cite{Reisinger-PoissonSpot2009}) has to be on the order of $\delta R\sim({R^2_\odot-\lambda r})^{1/2}-R_\odot\sim 10^4\lambda$ for an Arago spot to form at a distance of $r\sim100$~AU, which clearly is not the case. The Sun is not a perfect sphere due to its oblate shape and other deviations from spherical symmetry \cite{Park-etal:2017,Roxburgh:2001,Mecheri-2004}. In addition, the solar corona, which is composed of a free electron plasma, certainly is not stable on spatial scales of several hundreds of wavelengths (see \cite{Turyshev-Toth:2018-plasma} and references therein).  Thus, a study of the Arago spot in such conditions is of limited practical significance for the SGL (hence it was not carried out in \cite{Turyshev-Toth:2018}). Nevertheless, it may be relevant to some applications in astronomy and astrophysics, especially where longer wavelengths and strong gravitational fields are considered.

In Section~\ref{sec:shadow}, we have shown that there is no light in the shadow. In the case of the gravitational deflection of light, this shadow extends up to the heliocentric distance $R_\odot \leq r \leq z_0=R^2_\odot/2 r_g\simeq 547.8$~AU. In this Section, we consider the EM field in this region, which is characterized by Fresnel diffraction. Our objective is to determine the EM  field in the shadow of the Sun, especially on the optical axis of the SGL.

\subsection{The  Debye potential outside the shadow}
\label{sec:Deb-pot-outsdie}

The presence of the opaque Sun creates a spherical obscuration that blocks the rays of incident light with impact parameters,  $0\leq b\leq R^\star_\odot$. Thus, there are no light rays that can enter the shadow region. However, light rays that graze the Sun diffract on the edges of the solar disk into the shadow region, leading to the creation of the Arago spot on the optical axis of the SGL.

We begin by examining the expression (\ref{eq:P_g-tot+q}) together with (\ref{eq:Pi_ie*+f}). As these expressions suggest, the EM field outside the Sun is induced by the Debye potential that, to ${\cal O}(r_g^2)$ has the form:
{}
\begin{eqnarray}
\Pi (r, \theta)= -E_0
\frac{u}{2k^2r}\sum_{\ell=kR_\odot^*}^\infty i^{\ell}\frac{2\ell+1}{\ell(\ell+1)}e^{i\sigma_\ell} H^{(+)}_\ell(kr_g,kr) +E_0
\frac{u}{2k^2r}\sum_{\ell=1}^\infty i^{\ell}\frac{2\ell+1}{\ell(\ell+1)}e^{i\sigma_\ell}
H^{(-)}_\ell(kr_g,kr) P^{(1)}_\ell(\cos\theta).
  \label{eq:Pi_out}
\end{eqnarray}

The first sum in this expression is the remaining part of the outgoing wave that continues to move forward towards the larger values of $z$. The second wave is the incoming wave, which makes no contribution to the EM field on the optical axis in the area of interest (see \cite{Turyshev-Toth:2018}).  Therefore, we need to investigate the EM filed generated by the first term in (\ref{eq:Pi_out}), which we call $\Pi^{\rm out}_0 (r, \theta)$. Similarly to (\ref{eq:de_P-g}), we  use the asymptotic behavior of the Hankel functions (\ref{eq:FassC}) and present this potential in its asymptotic form as
\begin{eqnarray}
\Pi^{\rm out}_0 (r, \theta)&=&-E_0\frac{u}{2k^2r}e^{ik(r+r_g\ln 2kr)}\sum^{\infty}_{\ell=kR_\odot^*} \frac{2\ell+1}{\ell(\ell+1)}e^{i\big(2\sigma_\ell+\frac{\ell(\ell+1)}{2kr}\big)} P^{(1)}_\ell(\cos\theta)+{\cal O}(r_g^2).
\label{eq:de_P-g_out}
\end{eqnarray}

As we have shown in \cite{Turyshev-Toth:2017}, this is the potential that generates the EM field in the focal region of the SGL. Its components are given by  (\ref{eq:DB-sol00p}), together with $\alpha(r,\theta)$, $\beta(r,\theta)$ and $\gamma(r,\theta)$ from (\ref{eq:alpha})--(\ref{eq:gamma}) with the potential $\Pi$ replaced by $\Pi^{\rm out}_0 (r, \theta)$. Besides being responsible for the EM field in the interference region of the SGL (see Fig.~\ref{fig:regions}), this is the field that may be responsible for the light that is diffracted into the shadow to generate the Arago spot. To investigate this possibility, we need to compute the EM field generated by $\Pi^{\rm out}_0 (r, \theta)$. This can be done using  the same approach discussed in Sec.~\ref{sec:shadow}, where we studied the EM field due to the shadow potential $\delta \Pi^{\rm g} (r, \theta)$. To that extent, we take the $\Pi^{\rm out}_0 (r, \theta)$ from (\ref{eq:de_P-g_out}) and develop analytical expressions for the factors $\alpha(r,\theta), \beta(r,\theta)$ and $\gamma(\theta)$, which, to ${\cal O}\big(r^2_g\big)$, have the form
{}
\begin{eqnarray}
\alpha(r,\theta) &=& -E_0\frac{e^{ik(r+r_g\ln 2kr)}}{uk^2r^2}\sum^{\infty}_{\ell=kR_\odot^*}(\ell+{\textstyle\frac{1}{2}})e^{i\big(2\sigma_\ell+\frac{\ell(\ell+1)}{2kr}\big)}\Big\{u^2-\frac{ikr_g}{\ell(\ell+1)} \Big\}P^{(1)}_\ell(\cos\theta),
  \label{eq:alpha*1*o}\\
\beta(r,\theta) &=& E_0\frac{ue^{ik(r+r_g\ln 2kr)}}{ikr}\sum^{\infty}_{\ell=kR_\odot^*}\frac{\ell+{\textstyle\frac{1}{2}}}{\ell(\ell+1)}e^{i\big(2\sigma_\ell+\frac{\ell(\ell+1)}{2kr}\big)}
\Big\{\frac{\partial P^{(1)}_\ell(\cos\theta)}
{\partial \theta}\Big(1-\frac{\ell(\ell+1)}{2k^2r^2u^2}+\frac{ir_g}{2kr^2}\Big)+\frac{P^{(1)}_\ell(\cos\theta)}{\sin\theta}
 \Big\},
  \label{eq:beta*1*o}\\
\gamma(r,\theta) &=& E_0\frac{ue^{ik(r+r_g\ln 2kr)}}{ikr}\sum^{\infty}_{\ell=kR_\odot^*}\frac{\ell+{\textstyle\frac{1}{2}}}{\ell(\ell+1)}e^{i\big(2\sigma_\ell+\frac{\ell(\ell+1)}{2kr}\big)}
\Big\{\frac{\partial P^{(1)}_\ell(\cos\theta)}
{\partial \theta}+\frac{P^{(1)}_\ell(\cos\theta)}{\sin\theta}\Big(1-
\frac{\ell(\ell+1)}{2k^2r^2u^2}+\frac{ir_g}{2kr^2}\Big)
 \Big\}.~~~~~
  \label{eq:gamma*1*o}
\end{eqnarray}

We use these expressions to search for the light that may be diffracted by the sharp edges of a large circular obscuration creating an interference pattern on the optical axis in the presence of gravity.

\subsection{The function $\alpha(r,\theta)$ and the radial components of the EM field}
\label{sec:alpha-IF}

It is known that, in the absence of gravity, diffraction of light by an opaque sphere causes the appearance of the bright spot of Arago \cite{Harvey-Forgham:1984} within the Fresnel regime. We investigate if such a spot appears in the presence of gravity, which causes light rays to focus at the distance of  $R^2_\odot/2 r_g\simeq 547.8$~AU. The EM field in the shadow region is derived from the Debye potential (\ref{eq:P_g-tot}) and it is given by the factors $\alpha(r,\theta)$, $\beta(r,\theta)$ and $\gamma(r,\theta)$ from (\ref{eq:alpha*1*o})--(\ref{eq:gamma*1*o}).
We begin with the investigation of $\alpha(r,\theta)$, given by (\ref{eq:alpha*1*o}) as
{}
\begin{eqnarray}
\alpha(r,\theta) &=& -E_0\frac{e^{ik(r+r_g\ln 2kr)}}{k^2r^2}\sum_{\ell=kR^*_\odot}^{\infty}(\ell+{\textstyle\frac{1}{2}})\Big(u^2 -\frac{ikr_g}{\ell(\ell+1)} \Big)e^{i\big(2\sigma_\ell+\frac{\ell(\ell+1)}{2kr}\big)}P^{(1)}_\ell(\cos\theta).
  \label{eq:alpha*IF}
\end{eqnarray}

To evaluate expression (\ref{eq:alpha*IF}) directly on the optical axis or, more broadly, for $0\leq \theta \simeq \sqrt{2r_g/r}$, we use the asymptotic representation for $P^{(1)}_l(\cos\theta)$ from \cite{Bateman-Erdelyi:1953,Korn-Korn:1968,Kerker-book:1969}, valid when $\ell\to\infty$:
{}
\begin{eqnarray}
P^{(1)}_\ell(\cos\theta)&=& \frac{\ell+{\textstyle\frac{1}{2}}}{\cos{\textstyle\frac{1}{2}}\theta}J_1\big((\ell+{\textstyle\frac{1}{2}})2\sin{\textstyle\frac{1}{2}}\theta\big).
\label{eq:pi-l0}
\end{eqnarray}

This approximation may be used to transform (\ref{eq:alpha*IF}) as
{}
\begin{eqnarray}
\alpha(r,\theta) &=& -E_0\frac{e^{ik(r+r_g\ln 2kr)}}{k^2r^2\cos{\textstyle\frac{1}{2}}\theta}\sum_{\ell=kR_\odot^*}^{\infty}(\ell+{\textstyle\frac{1}{2}})^2\Big(u^2-\frac{ikr_g}{\ell(\ell+1)} \Big)e^{i\big(2\sigma_\ell+\frac{\ell(\ell+1)}{2kr}\big)}
J_1\big((\ell+{\textstyle\frac{1}{2}})2\sin{\textstyle\frac{1}{2}}\theta\big).
  \label{eq:alpha*IF*}
\end{eqnarray}

It is easy to see that for $\theta=0$, $\alpha(r,\theta)$ vanishes. Therefore,  even in the absence of gravity, when $r_g=0$, the radial components of the EM field, $(D_r, B_r)$, cannot be responsible for the bright spot of Arago \cite{Harvey-Forgham:1984,Cash:2011}. This remains true when the propagation of light is affected by gravity, when $r_g\not=0$. Therefore, we turn our attention to the other two components of the EM field, governed by the functions $\beta(r,\theta)$ and $\gamma(r,\theta)$.

\subsection{The functions $\beta(r,\theta)$ and $\gamma(r,\theta)$
and the spot of Arago in the presence of gravity}
\label{sec:beta-IF}

In studying the solar shadow region, we consider rays with impact parameters $0\lesssim b\leq R^\star_\odot$, which correspond to the partial momenta  $1\lesssim \ell \leq kR^\star_\odot$. We are concerned with the interaction  of light with the surface of the sphere where $b=R^\star_\odot$, which corresponds to $\ell_{\rm 0}=k  R^\star_\odot$. In this case, we may neglect terms $\propto \ell^2/4k^2r^2u^2\simeq (R^*/2r)^2$ in (\ref{eq:beta*1*o})--(\ref{eq:gamma*1*o}), as their contributions are very small, reaching $5.4\times 10^{-6}$ at $r=1$~AU.  In addition, we may neglect the terms $ir_g/2kr^2$ in the functions $\beta(r,\theta)$ and $\gamma(r,\theta)$ from (\ref{eq:beta*1*o}) and (\ref{eq:gamma*1*o}), correspondingly. This allows us to present these functions as
{}
\begin{eqnarray}
\beta(r,\theta) =\gamma(r,\theta)\equiv {\cal A}(r,\theta),
  \label{eq:beta-IF0*}
\end{eqnarray}
where the function ${\cal A}(r,\theta)$ is given as
{}
\begin{eqnarray}
{\cal A}(r,\theta)&=& E_0\frac{e^{ik(r+r_g\ln 2kr)}}{ikr}\sum_{\ell=kR^*_\odot}^\infty\frac{\ell+{\textstyle\frac{1}{2}}}{\ell(\ell+1)}e^{i\big(2\sigma_\ell+\frac{\ell(\ell+1)}{2kr}\big)}
\Big\{\frac{\partial P^{(1)}_\ell(\cos\theta)}
{\partial \theta}+\frac{P^{(1)}_\ell(\cos\theta)}{\sin\theta}
 \Big\}.
  \label{eq:S-rt}
\end{eqnarray}

To evaluate the magnitude of this function, we  need to use the asymptotic behavior of the Legendre-polynomials $P^{(1)}_{l}(\cos\theta)$ in the relevant regime \cite{vandeHulst-book-1981}, characterized by $w=(\ell+{\textstyle\frac{1}{2}})\theta$ being fixed and $\ell\rightarrow\infty$. These are given in the following form:
{}
\begin{eqnarray}
\frac{P^{(1)}_\ell(\cos\theta)}{\sin\theta}&=& {\textstyle\frac{1}{2}}\ell(\ell+1)\Big(J_0(w)+J_2(w)\Big),
\qquad
\frac{dP^{(1)}_\ell(\cos\theta)}{d\theta}= {\textstyle\frac{1}{2}}\ell(\ell+1)\Big(J_0(w)-J_2(w)\Big).
\label{eq:tau-l}
\end{eqnarray}

Using the expressions (\ref{eq:tau-l}), we transform (\ref{eq:S-rt}) as follows:
{}
\begin{eqnarray}
{\cal A}(r,\theta)&=& E_0\frac{e^{ik(r+r_g\ln 2kr)}}{ikr}\sum_{\ell=kR^*_\odot}^\infty (\ell+{\textstyle\frac{1}{2}})e^{i\big(2\sigma_\ell+\frac{\ell(\ell+1)}{2kr}\big)}J_0\big((\ell+{\textstyle\frac{1}{2}})\theta\big).
  \label{eq:S-rt+}
\end{eqnarray}
Thus, we see that $\beta(r,\theta) $ and $\gamma(r,\theta)$  may be responsible for the bright spot of Arago, as neither of them vanish at $\theta=0$. By replacing the sum in (\ref{eq:S-rt+}) with an integral, we have:
{}
\begin{eqnarray}
{\cal A}(r,\theta)&=& E_0\frac{e^{ik(r+r_g\ln 2kr)}}{ikr}\int_{\ell=kR^*_\odot}^\infty e^{i\big(2\sigma_\ell+\frac{\ell^2}{2kr}\big)}J_0\big(\ell\theta\big)\ell d\ell.  \label{eq:S-rt*2}
\end{eqnarray}

The analysis of the EM field in the interference region of the SGL is based on the investigation of the refractive properties of the gravitational field of the Sun. It tells nothing about diffraction of light by the spherical obscuration in the presence of gravity. However, the same integral (\ref{eq:S-rt+}) also describes the diffraction of light on the boundaries of a spherical obscuration. Evaluation of this integral for these purposes requires the tools of  numerical analysis, as no analytical solution to (\ref{eq:S-rt+}) is known. For this purpose, we present a set of useful approximations. We first examine the total phase of this expression, $\varphi$, which, by using expression for $\sigma_\ell$ from (\ref{eq:sig-l*}) and taking $\ell\simeq kb$,
may be given as
{}
\begin{eqnarray}
\varphi&=&
kr+\frac{\ell^2}{2kr}-kr_g\ln \frac{\ell^2}{2kr}\simeq
k\Big(r+\frac{b^2}{2r}-r_g\ln \frac{kb^2}{2r}\Big)+{\cal O}(r_g^2).
\label{eq:ph-diff}
\end{eqnarray}
The terms in this expression have clear meaning: The first one is the classical phase due to a free-wave propagation towards the image plane.  The second term is the phase of the spherical wave obeying the Fresnel approximation, which is used to describe the bright spot of Arago. The third term is new and is  due to gravitational Shapiro delay.

Evaluating expression (\ref{eq:ph-diff}) numerically for $b=R_\odot$, we see that within the shadow region, both of these terms are about the same order of magnitude. However, at  heliocentric distances $r<r_0\simeq20.3$~AU, the phase (\ref{eq:ph-diff}) is dominated by the first term due to the classical Fresnel diffraction. But then,  after crossing $r_0$, the phase changes sign and, for  $r>20.3$~AU, it becomes increasingly dominated by the second term due to the refraction of light in gravity. Such a dominant behavior is because of the  gravitational $1/r$ potential (\ref{eq:w-PN}) whose presence is evident even at very large distances from the gravitating sphere.

With result (\ref{eq:ph-diff}) we present (\ref{eq:S-rt*2}) in the following identical form:
{}
\begin{eqnarray}
{\cal A}(r,\theta)&=& E_0\frac{e^{ik}}{ikr}\int_{\ell=kR^*_\odot}^\infty e^{i\big(\frac{\ell^2}{2kr}-kr_g\ln \frac{\ell^2}{2kr}\big)}J_0\big(\ell\theta\big)\ell d\ell,
\label{eq:S-rt*2-di}
\end{eqnarray}
where all three phase contributions,  namely the phase shift due to the length of the propagation path, the phase shift due to Fresnel diffraction, and the Shapiro phase shift due to gravity are clearly shown.

We first consider  (\ref{eq:S-rt*2-di}) in flat spacetime, i.e., by setting  $r_g=0$, and evaluating the resulting integral by parts:
{}
\begin{eqnarray}
{\cal A}_0(r,\theta)&=& E_0\frac{e^{ikr}}{ikr}\int_{\ell=kR^*_\odot}^\infty e^{i\frac{\ell^2}{2kr}}J_0\big(\ell\theta\big)  \ell d\ell =
E_0e^{ikr} \Big\{e^{i\frac{k R^{\star2}_\odot}{2r}}J_0\big(k R^\star_\odot \theta\big)-\theta\int_{\ell=kR^*_\odot}^\infty e^{i\frac{\ell^2}{2kr}}J_1\big(\ell\theta\big) d\ell \Big\}.
\label{eq:S-rt*2-0+s}
\end{eqnarray}
Recognizing that for $\ell\theta\rightarrow0$, the Bessel function $J_1(\ell \theta)$ behaves as $J_1(\ell \theta)={\textstyle\frac{1}{2}}\ell \theta-{\textstyle\frac{1}{16}}\ell^3 \theta^3+{\cal O}(\ell^5 \theta^5)$, we keep only the leading term for $J_1(\ell \theta)$ and present the second term on the right hand side of (\ref{eq:S-rt*2-0+s}) as below:
{}
\begin{eqnarray}
\theta\int_{\ell=kR^*_\odot}^\infty e^{i\frac{\ell^2}{2kr}}J_1\big(\ell\theta\big) d\ell \simeq {\textstyle\frac{1}{2}}\theta^2 \int_{\ell=kR^*_\odot}^\infty e^{i\frac{\ell^2}{2kr}} \ell d\ell \simeq -i\frac{k\rho^2}{2r}  e^{i\frac{\ell^2}{2kr}}  \Big|_{\ell=kR^*_\odot}^\infty,
\label{eq:S-rt*2-01}
\end{eqnarray}
where we used $\theta=\rho/r$ and, for small angles, $r\simeq z$, with $z$ being the heliocentric distance to the image plane and $\rho$ being the distance from the optical axis on that plane. Clearly, (\ref{eq:S-rt*2-01}) vanishes both when $r$ increases and/or when $\rho\rightarrow 0$, approaching the optical axis.  Therefore, this term may be neglected.  As  a result, (\ref{eq:S-rt*2-0+s}) takes the form
{}
\begin{eqnarray}
{\cal A}_0(r,\theta)&=& E_0\exp\Big[ik\Big(r+\frac{R^{\star2}_\odot}{2r}\Big)\Big]J_0\big(k R^\star_\odot \theta\big),
\label{eq:S-rt*2-0}
\end{eqnarray}
which has the properties of the bright spot of Arago \cite{Harvey-Forgham:1984}, describing the corresponding intensity distribution on the image plane.  Thus, we have established that our wave-optical treatment recovers the well-known features of diffraction by the boundary of a spherical obscuration.

The same result may be obtained by using the approach relying on the scattering transfer function, $\eta(\ell)$, that captures the interaction of light with the sphere. The normalized scattering function in this case has the from
$\eta(\ell)=\delta(\ell-kR^\star_\odot)ir/R^\star_\odot$ (which is consistent with \cite{Greider-Glassgold:1960,Frahn-Venter:1963}, which deals with the problems of nuclear scattering, and \cite{Aime:2013} that deals with practical applications of the Arago spot in designing solar coronagraphs), where $\delta(\cdot)$ is Dirac's delta function, can be used to select the innermost impact parameter just outside the Sun, representing rays of light that are diffracted by the edge of the solar disk. Using this scattering function, we perform the integration of (\ref{eq:S-rt*2-0+s})  as
{}
\begin{eqnarray}
{\cal A}(r,\theta)&=& E_0\frac{e^{ikr}}{ikr}\int_{0}^{\infty} \eta(\ell)e^{i\frac{\ell^2}{2kr}}J_0\big(\ell\theta\big)\ell d\ell =E_0e^{ikr} e^{i\frac{k R^{\star2}_\odot}{2r}}J_0\big(k R^\star_\odot \theta\big).
  \label{eq:S-rt*2-02}
\end{eqnarray}
Clearly, the result (\ref{eq:S-rt*2-02}) is identical to (\ref{eq:S-rt*2-0}).
Using this expression for the functions $\beta(r,\theta)$ and $\gamma(r,\theta)$ in (\ref{eq:DB-sol00p}), we compute the components of the corresponding EM field and verify that the Poynting vector of the EM field that is responsible for the Arago spot has only one non-vanishing component, namely that along the $z$-axis:
{}
\begin{eqnarray}
S^{\rm Arago}_z&=& \frac{c}{8\pi}E^2_0J^2_0\Big(2\pi \frac{\rho}{\lambda}\frac{R^\star_\odot}{z}\Big).
  \label{eq:S-rt*3}
\end{eqnarray}
  The light intensity distribution on the image plane resulted from (\ref{eq:S-rt*3}) is consistent with that of the bright spot of Arago (see  \cite{Harvey-Forgham:1984,Cash:2011,Aime:2013} and references therein), thus confirming the know result \cite{Born-Wolf:1999}.

Now we consider the same process, but in the presence of gravity. Unfortunately, it is not possible to analytically integrate  (\ref{eq:S-rt*2-di}) in a general case for  $r_g\not=0$; however, we can easily do this for $\theta=0$, which is sufficient for our purposes. Indeed,  using the properties of the upper incomplete Gamma function,  $\Gamma[a, z]$,  \cite{Abramovitz-Stegun:1965}.
We  integrate (\ref{eq:S-rt*2-di}) for  $\theta=0$ as
{}
\begin{eqnarray}
{\cal A}(r,0)&=& E_0e^{ikr}\Big\{
\exp\Big[ik\Big(\frac{R^{*2}_\odot}{2r}-r_g\ln \frac{kR^{*2}_\odot}{2r}\Big)\Big]-e^{\frac{\pi}{2} kr_g}\Big(\Gamma\big[1-ikr_g,-i\frac{\infty^2}{2kr}\big]+ikr_g\Gamma\big[-ikr_g,-i\frac{kR^{*2}_\odot}{2r}\big]\Big)\Big\}.~~~
\label{eq:S-rt*2+=}
\end{eqnarray}
Given realistic values of the parameters involved (i.e., $k$, $r_g$ and $R_\odot$), the part containing the incomplete Gamma function vanishes. As a result,  (\ref{eq:S-rt*2+=}) takes the form:
{}
\begin{eqnarray}
{\cal A}(r,0)&=& E_0\exp\Big[ik\Big(r+\frac{R^{*2}_\odot}{2r}-r_g\ln \frac{kR^{*2}_\odot}{2r}\Big)\Big].
\label{eq:S-rt*2+=F}
\end{eqnarray}
Clearly, in the limit $r_g\rightarrow0$, the result (\ref{eq:S-rt*2+=F}) reduces to (\ref{eq:S-rt*2-0}), taken on-axis. Thus, we see that apart from an additional gravity-induced phase shift, the Arago spot appears, as expected, in the shadow region.
To integrate (\ref{eq:S-rt*2-di}) in the general case, we may use the approach relying on the scattering function $\eta(\ell)$ that was discussed above, which yields
{}
\begin{eqnarray}
{\cal A}(r,\theta)&=& E_0\frac{e^{ik}}{ikr}\int_{0}^\infty \eta(\ell) e^{i\big(\frac{\ell^2}{2kr}-kr_g\ln \frac{\ell^2}{2kr}\big)}J_0\big(\ell\theta\big)\ell d\ell=
E_0\exp\Big[ik\Big(r+\frac{R^{*2}_\odot}{2r}-r_g\ln \frac{kR^{*2}_\odot}{2r}\Big)\Big]J_0\big(kR^*_\odot \theta\big).
\label{eq:S-rt*2+g+}
\end{eqnarray}
Thus, although in the presence of gravity, the Arago spot acquires an additional large phase shift, its magnitude remains unchanged. The intensity distribution of the EM field in the shadow on the optical axis behind the large opaque spherical obscuration is still given by (\ref{eq:S-rt*3}). These results consistent with the approach relying on the sharp boundary of an opaque sphere. A more refined modeling of the interaction of the EM wave with the surface of the sphere is possible using methods discussed in \cite{Greider-Glassgold:1960,vandeHulst-book-1981,Grandy-book-2000}. However, this is beyond the scope of this paper.

For completeness, we mention that the EM field in the interference region of the SGL, for $r\geq R^2_\odot/2r_g=547.8$~AU, was studied in  \cite{Turyshev-Toth:2017} (see, (121) therein). In that case, the integral (\ref{eq:S-rt*2-di}) was evaluated by using the method of stationary phase with the functions  $\beta(r,\theta)=\gamma(r,\theta)\equiv {\cal A}_{\tt SGL}(r,\theta)$ were obtained in the form
{}
\begin{eqnarray}
{\cal A}_{\tt SGL}(r,\theta)&=& E_0\Big({\frac{2\pi kr_g}{1-e^{-2\pi kr_g}}}\Big)^\frac{1}{2} J_0\Big(k\sqrt{2r_gr} \theta\Big)e^{ikr}.
  \label{eq:S-rt*2inc}
\end{eqnarray}
This expression describes the properties of the only non-vanishing component of the Poynting vector along the $z$-axis, ${S}_z$, which, from (\ref{eq:S-rt*2inc}), together with $k=2\pi/\lambda$ and $\theta=\rho/z$, is  given as (see \cite{Turyshev-Toth:2017} for details)
{}
\begin{eqnarray}
{S}_z&=&\frac{c}{8\pi}E_0^2\frac{4\pi^2}{1-e^{-4\pi^2 r_g/\lambda}}\frac{r_g}{\lambda}\, J^2_0\Big(2\pi\frac{\rho}{\lambda}\sqrt{\frac{2r_g}{z}}\Big).
\label{eq:S_z*6z}
\end{eqnarray}
This expression summarizes the optical properties of the SGL in its interference region providing the means to describe its major signal amplification and  significant angular resolution \cite{Turyshev-Toth:2017}.
As evident from (\ref{eq:S_z*6z}), the largest amplification of the SGL occurs on the $z$ axis (i.e., with $\rho=0$), where no other fields are present. This is a somewhat  intuitive result, but, nevertheless, now we have provided a confirmation for this fact relying on the  wave-optical treatment.

\section{Discussion and Conclusions}
\label{sec:disc}

We investigated the EM field in the shadow cast by a large gravitating sphere and developed a wave-optical treatment of this problem.  The results obtained here are relevant to our ongoing work on the imaging with the SGL \cite{Turyshev-Toth:2017,Turyshev-etal:2018,Turyshev-etal:2018-wp}. These results are also relevant to some other practical applications, such as Rutherford scattering microscopy \cite{Selmke:2015} where one encounters similar potentials in addressing the scattering of focused beams in a refractive medium. Thus, description of the shadow in the case of a repulsive Coulomb potential (i.e., the gravitational field produced by a relativistic mass-monopole (\ref{eq:w-PN})) presented here has an immediate practical value in describing the beam diffraction pattern for light as well as other massless particles.

However, our primary motivation is the study of the optical properties of the SGL and especially its potential for high-resolution imaging and spectroscopy of exoplanets. In this regard, we have demonstrated that the fully absorbing boundary conditions introduced in Sec.~\ref{sec:em-waves-gr+pl} allow for a proper description of the shadow region behind the Sun. We have shown that, in the presence of gravity, light rays are bent towards the Sun and the resulting solar shadow has the shape of the rotational hyperboloid. We have also shown that there is no light in the shadow region, except for that diffracted by the sharp boundary of a large spherical obscuration to form the bright spot of Arago within the shadow of the SGL on its optical axis. Furthermore, in the presence of gravity, the total EM field in the interference region is given solely by the incident light, which is deflected by the solar gravity towards the optical axis where it forms a caustic, thereby significantly amplifying the intensity of the incident light, as discussed in \cite{Turyshev-Toth:2017}.

Our next steps involve investigation of the effect of the solar plasma (using, for instance, the approach developed in \cite{Turyshev-Toth:2018-plasma}) and of the solar non-sphericity and optical properties of  the caustic formed by the gravitational field of the extended Sun. These topics are currently being investigated; results, once available,  will be reported elsewhere.

\begin{acknowledgments}
This work in part was performed at the Jet Propulsion Laboratory, California Institute of Technology, under a contract with the National Aeronautics and Space Administration.

\end{acknowledgments}


\end{document}